\documentclass{aa}  

\usepackage[colorlinks, allcolors=blue]{hyperref}
\usepackage{amsmath} 

\usepackage{graphicx}
\usepackage{xcolor}
\usepackage{txfonts}
\usepackage{longtable}
\usepackage{geometry}
\usepackage{datetime}
\usepackage{lscape}
\usepackage{float}
\usepackage{placeins}
\newcommand{\tablecomments}[1]{%
    \par\vskip1pt
    \noindent
    \vrule height 11pt depth 2pt width 0pt
    \textsc{Note} -- %
    #1\par
    \vskip1pt
}

\begin{document} 

\title{X-ray spectral properties of the accreting millisecond pulsar IGR J17498$-$2921 during its 2023 outburst}

\author{G.~Illiano\inst{1,2,3}, A.~Papitto \inst{1}, 
A.~Marino \inst{4,5}, 
T. E. Strohmayer \inst{6,7},
A. Sanna \inst{8},
T. Di Salvo \inst{9},
R. La Placa\inst{1},
F. Ambrosino\inst{1},
A. Miraval Zanon\inst{10},
F. Coti Zelati \inst{4,5,11},
C. Ballocco\inst{3,1},
C. Malacaria\inst{1},
A.~Ghedina\inst{12},
M.~Cecconi\inst{12},
M.~Gonzales\inst{12},
F.~Leone\inst{13,14}
}

\institute{INAF-Osservatorio Astronomico di Roma, Via Frascati 33, I-00076, Monte Porzio Catone (RM), Italy\\
\email{giulia.illiano@inaf.it}
\and Tor Vergata University of Rome, Via della Ricerca Scientifica 1, I-00133 Roma, Italy
\and Sapienza Università di Roma, Piazzale Aldo Moro 5, I-00185 Rome, Italy
\and Institute of Space Sciences (ICE, CSIC), Campus UAB, Carrer de Can Magrans s/n, E-08193 Barcelona, Spain
\and Institut d’Estudis Espacials de Catalunya (IEEC), E-08860 Castelldefels (Barcelona), Spain 
\and Astrophysics Science Division, NASA Goddard Space Flight Center, Greenbelt, MD 20771, USA
\and Joint Space-Science Institute, NASA Goddard Space Flight Center, Greenbelt, MD 20771, USA
\and Dipartimento di Fisica, Università degli Studi di Cagliari, SP Monserrato-Sestu km 0.7, I-09042 Monserrato, Italy
\and Università degli Studi di Palermo, Dipartimento di Fisica e Chimica, via Archirafi 36, 90123 Palermo, Italy
\and ASI - Agenzia Spaziale Italiana, Via del Politecnico snc, 00133 Roma, Italy
\and INAF–Osservatorio Astronomico di Brera, Via Bianchi 46, I-23807 Merate (LC), Italy
\and Fundación Galileo Galilei - INAF, La Palma, Spain
\and Dipartimento di Fisica e Astronomia, Sezione Astrofisica, Universita` di Catania, Via S. Sofia 78, I-95123 Catania, Italy 
\and INAF – Osservatorio Astrofisico di Catania, Via S. Sofia 78, I-95123 Catania, Italy
}             

\date{}
\authorrunning{Illiano et al.}
\titlerunning{X-ray spectral properties of the accreting millisecond pulsar IGR J17498$-$2921 during its 2023 outburst} 
 
     \abstract
    {We present a comprehensive study of the X-ray spectral properties of the accreting millisecond pulsar IGR J17498$-$2921 during its 2023 outburst. Similar to other accreting millisecond pulsars, the broad-band spectral emission observed quasi-simultaneously by NICER and NuSTAR is well described by an absorbed Comptonized emission with an electron temperature of $\sim$17 keV plus a disk reflection component. The broadening of the disk reflection spectral features, such as a prominent iron emission line at 6.4--6.7~keV, is consistent with the relativistic motion of matter in a disk truncated at $\sim$$21 \, \mathrm{R_g}$ from the source, near the Keplerian co-rotation radius. From the high-cadence monitoring data obtained with NICER, we observe that the evolution of the photon index and the temperature of seed photons tracks variations in the X-ray flux. This is particularly evident close to a sudden $\sim$-0.25 cycles jump in the pulse phase, which occurs immediately following an X-ray flux flare and a drop in the pulse amplitude below the $3\sigma$ detection threshold. We also report on the non-detection of optical pulsations with TNG/SiFAP2 from the highly absorbed optical counterpart.}


   \maketitle
%

\section{Introduction}
Accreting millisecond pulsars (AMSPs; see, e.g., \citealt{DiSalvo_Sanna_2022ASSL, Patruno_Watts_2021ASSL, Campana_DiSalvo_2018ASSL}) are relatively low-magnetized ($B \sim 10^{8}-10^{9} \, \mathrm{G}$), rapidly rotating ($P_{\textrm{spin}} \leq 10 \, \mathrm{ms}$) neutron stars (NSs) usually hosted in tight binary systems. They attain their fast spin periods through a Gyr-long phase in which they shine as bright low-mass X-ray binaries (LMXBs), fueled by accreting matter and angular momentum transfer from a sub-solar donor star. When accretion stops, the NSs turn on as rotation-powered MSPs, making them descendants of AMSPs \citep{Alpar_1982Natur, Radhakrishnan_Srinivasan_1982CSci}. The first direct evidence of fast X-ray pulsations from an accretion-powered neutron star in a LMXB came in 1998, proving that spin up to millisecond periods through accretion does occur in such systems \citep{Wijnands_VanDerKlis_1998Natur}.
AMSPs stand out from the broader class of LMXBs for their coherent pulsations at the NS spin period, which serve as critical diagnostic tools to measure the NS properties and understand the emission mechanisms in different states.
These sources spend most of their life in quiescence, featuring X-ray luminosity of $L_X \lesssim 10^{31}-10^{33} \, \mathrm{erg \, s^{-1}}$ \citep[e.g.,][]{Campana_DiSalvo_2018ASSL}. With recurrence times that typically vary from a few months to years, they exhibit outburst phases, during which their X-ray luminosity can increase up to $L_X \sim 10^{36}-10^{37} \, \mathrm{erg \, s^{-1}}$.
At the time of writing, 26\footnote{This count includes the three confirmed transitional millisecond pulsars (tMSPs), which are MSPs observed to swing between accretion-powered and rotation-powered states within a few days \citep[see, e.g.,][]{Papitto_DeMartino_2022ASSL}. However, to date, only the tMSP IGR J18245$-$2452 has exhibited an accretion outburst with a peak X-ray luminosity and duration comparable to other AMSPs \citep{Papitto_2013Natur}.} AMSPs have been discovered, with the most recent one identified in February 2024 \citep{Molkov_2024arXiv, Ng_2024}.

Unlike the typical behavior observed in non-pulsating LMXBs, the majority of AMSPs do not undergo transitions between hard and soft states during outbursts, leading to their common classification as hard X-ray transients \citep{DiSalvo_Sanna_2022ASSL}.
Their X-ray spectra are usually well described by a Comptonization component featuring an electron temperature in the tens of keV range, alongside one or two black body components, representing thermal emission from a colder accretion disk plus a hotter emission region from a boundary layer between the disk and the NS surface, or directly from the surface itself \citep{Gierlinski_2002MNRAS, Gierlinski_2005MNRAS, Poutanen_2006AdSpR, Marino_2023MNRAS}.
Additionally, a fraction of Comptonized photons may illuminate the disk, resulting in a reflection component \citep[see, e.g.,][]{Papitto_2010MNRAS, DiSalvo_2019MNRAS, Sharma_2019MNRAS}. The reflection process often manifests as a broad iron K$\alpha$ line at energies around 6.4 – 6.7 keV, and a Compton back-scattering hump at $\sim$$10-30$ keV \citep{Gierlinski_2002MNRAS, DiSalvo_Sanna_2022ASSL}.
Accurate modeling of these spectral features provides valuable insights into various physical parameters, including the ionization state and the truncation radius of the inner accretion disk, the system's inclination, and the fraction of reflected Comptonized photons.

IGR J17498$-$2921 was discovered during an outburst in August 2011 by \textrm{INTEGRAL} \citep{Gibaud_2011ATel}. Coherent 401 Hz X-ray pulsations were detected with \textrm{RXTE}, identifying the X-ray transient as an AMSP \citep{Papitto_2011A&A}.
A broad-band spectral analysis during the 2011 outburst was performed by \citet{Falanga_2012A&A}, revealing spectral stability throughout the outburst and X-ray emission described by thermal Comptonization originating from a heated slab located above the NS surface. Subsequent observations revealed thermonuclear X-ray bursts and burst oscillations \citep{Ferrigno_2011ATel, Chakraborty_2012MNRAS}, leading to a distance estimate of $\sim$7.6 kpc from a photospheric radius expansion burst observed with \textrm{RXTE} \citep{Linares_2011ATel}. The NS orbits a $> 0.17 \, \mathrm{M_\odot}$ companion, with an orbital period of $\sim$3.8 hr \citep{Papitto_2011A&A}. The source transitioned back to quiescence at the end of September 2011 \citep{Linares_2011ATel_quiescence}.
On 2023 April 13-15, \textrm{INTEGRAL} identified a potential new outburst from IGR J17498$-$2921 \citep{INTEGRAL_2023ATel}, later confirmed by prompt, targeted follow-up \textrm{NICER} observations \citep{Sanna_2023ATel_inizio_outburst}. \textrm{NICER} detected the source at a count rate of $\sim$90 c/s in the 0.5-10 keV band, well above the $\sim$1 c/s background level.\\

In this paper, we report on observations of IGR J17498$-$2921 during its 2023 outburst performed with the X-ray telescopes NICER and NuSTAR, and the fast optical photometer SiFAP2 mounted at the INAF Telescopio Nazionale Galileo (TNG). In Sect. \ref{sec:obs}, we describe NICER, NuSTAR, and SiFAP2 data and their reduction. Section \ref{sec:spectal_analysis} is dedicated to the spectral analysis, discussing both the broad-band spectrum and the spectral evolution throughout the outburst. We present a correlated spectral--timing analysis in Sect. \ref{sec:specral-timing_evolution}, with a more detailed study to be presented in \citet{Strohmayer_in_prep}, and the upper limit on the optical pulsations in Sect. \ref{sec:optical_pulsations}. We discuss our findings in Sect. \ref{sec:discussion}.

\section{Observations} \label{sec:obs}

\subsection{NICER} \label{sec:NICER}
\begin{figure*}
   \centering
   \includegraphics[width=17cm]{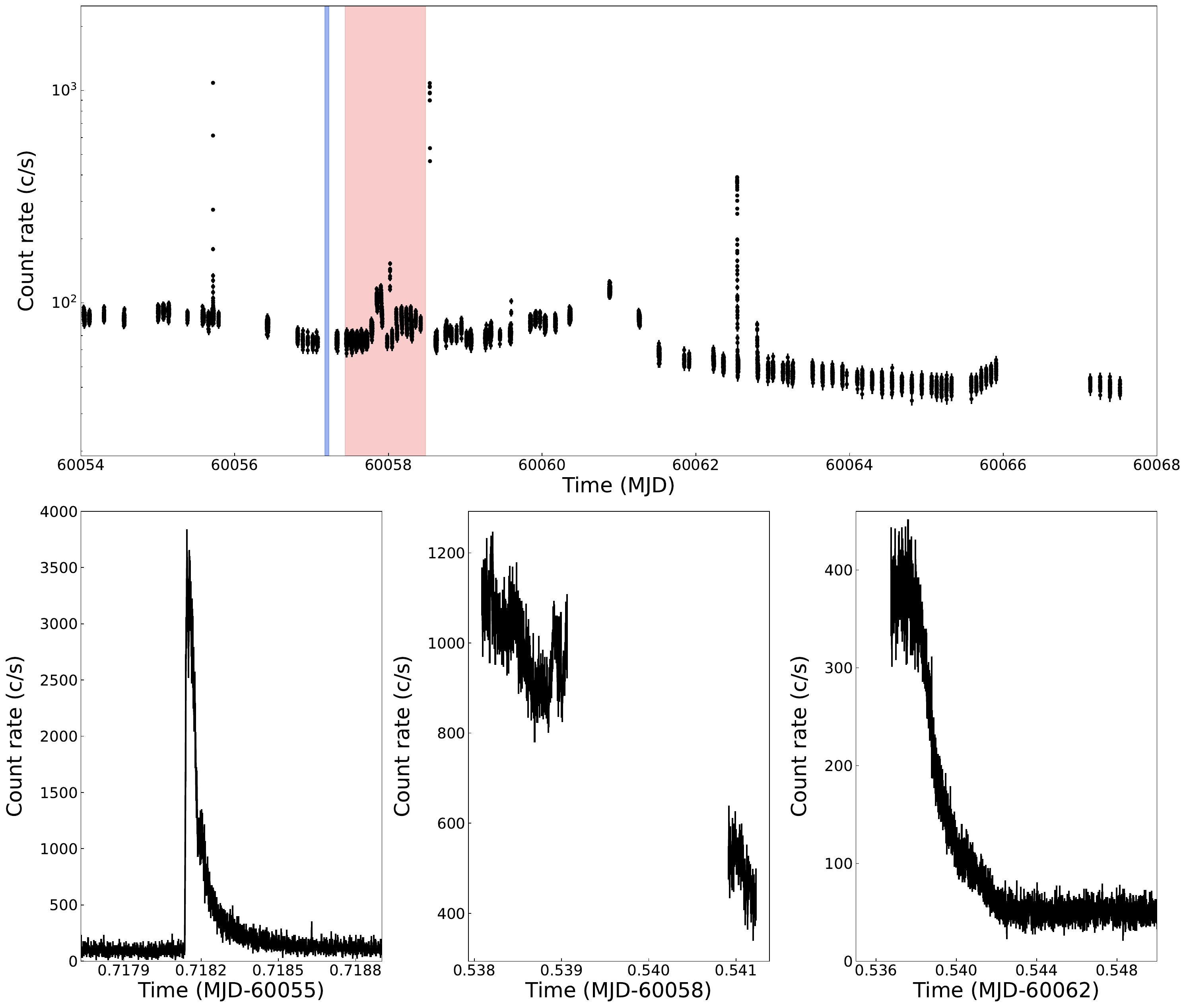}
      \caption{\textit{Top panel}: NICER 0.5–10 keV light curve using 10-s bins. The blue-shaded region represents the epoch of SiFAP2 observation, while the red-shaded region represents NuSTAR observation.
      \textit{Bottom left panel}: Type I X-ray burst in the 0.5-10 keV light curve of ObsID 6560010101 binned at 0.1 seconds. 
      \textit{Bottom middle panel}: Zoom-in of the light curve of ObsID 6560010104 binned at 0.5 seconds showing the increase of the X-ray flux just after the end of NuSTAR observation. 
      \textit{Bottom right panel}: Tail of type-I X-ray burst in the 0.5-10 keV light curve of ObsID 6560010108 binned at 1 second.
      }\label{Fig:lc_and_bursts}
\end{figure*}

The NASA soft X-ray telescope Neutron star Interior Composition Explorer (\textrm{NICER}; \citealt{NICER_Gendreau_2012}) monitored IGR J17498$-$2921 from 2023 April 20 (MJD 60054) until 2023 July 8 (MJD 60133; ObsIDs starting with 620377 and 656001; total exposure of $\sim$130 ks; PI: P. Bult).
We analyzed NICER observations using HEASoft version 6.33.2 and NICERDAS version 12, along with the calibration database (CALDB) version 20240206. After retrieving the latest geomagnetic data with the \texttt{nigeodown} tool, we processed the observational data using \texttt{nicerl2} task. We corrected the photon arrival times to the Solar System Barycenter using the JPL ephemerides DE405 \citep{Standish_DE405}. We adopted the source coordinates R.A. (J2000)=$17^{\mathrm{h}}49^{\mathrm{m}}55\fs35$ and DEC. (J2000)=$-29\degr19'19\farcs6$ \citep{Chakrabarty_2011ATel}.
We used the same parameters to correct arrival times observed by NuSTAR and TNG/SiFAP2.
The obtained cleaned event files were further used to generate spectral products using the \texttt{nicerl3-spect} task, with \texttt{SCORPEON} background model version 22 setting the options
\texttt{bkgformat=file}. 
Using the ftools \texttt{ftgrouppha}, automatically called by the \texttt{nicerl3-spect} task, we grouped the spectra using the ``optimal binning'' scheme \citep{Kaastra_Bleeker_2016A&A}, with a minimum of 25 counts per energy bin (using the options \texttt{GROUPTYPE=OPTMIN} and \texttt{GROUPSCALE = 25}). 
We extracted the light curves in the 0.5–10 keV energy range with different time bins using the \texttt{nicerl3-lc} pipeline, which
estimates the background level using the \texttt{SCORPEON}\footnote{See \url{https://heasarc.gsfc.nasa.gov/docs/nicer/analysis_threads/scorpeon-overview/}} background model (Fig. \ref{Fig:lc_and_bursts}).
To perform the spectral and timing analysis of the persistent emission, we filtered out the two type-I X-ray bursts observed during the NICER coverage (see the bottom left and right panels of Fig. \ref{Fig:lc_and_bursts}). We estimated the bursts' durations by computing the mean, $\mu$, and the standard deviation, $\sigma$, of the pre-burst counts' distribution in the light curves binned at 0.1 s for the first burst and 1 s for the second one due to the lower count rate. The end of the burst was identified as the point in the burst tail where the counts dropped below $\mu$ + 2$\sigma$. The excluded time intervals were (60055.718125--60055.718486) MJD, and (60062.536726--60062.545164) MJD.

\subsection{NuSTAR}
The NASA hard X-ray telescope Nuclear Spectroscopic Telescope Array (NuSTAR; \citealt{Harrison_2013ApJ}) observed IGR J17498$-$2921 on 2023 April 23 for a net exposure of $\sim$41 ks (ObsID 90901317002).
We processed and analyzed the data using NuSTAR data analysis package (\texttt{NUSTARDAS}) version v2.1.2 with the CALDB 20240325.
We selected source and background events from circular regions of $80^{\prime \prime}$ radius centered at the source location and in a source-free region away from the source, respectively.
We extracted the spectra for the two focal plane modules (FPMA and FPMB) and the corresponding response files using the \texttt{nuproducts} command.
During the NuSTAR observation, two type-I bursts were detected. The first burst was filtered out following the procedure outlined in Sect. \ref{sec:NICER}, with the excluded time interval specified as (60057.804093 -- 60057.804753) MJD. The second burst occurred when the satellite was in proximity to the South Atlantic Anomaly (SAA) and was automatically removed by selecting the \texttt{SSAMODE=OPTIMIZED} option in the  \texttt{nupipeline} task.

\subsection{TNG/SiFAP2}
The fast optical photometer SiFAP2 \citep{Ghedina_2018SPIE}, mounted at the 3.58-meter INAF Telescopio Nazionale Galileo (TNG; \citealt{Barbieri_1994SPIE}), observed IGR J17498$-$2921 on 2023 April 23 starting on 04:04:30 UTC for $\sim$4.5 ks.
We employed a white ﬁlter spanning the 320–900 nm band and peaking between 400 and 600 nm (i.e., roughly corresponding to the B and V Johnson ﬁlters; see Supplementary Figure 1 in \citealt{Ambrosino17}).
We used a proprietary pipeline for the data reduction.
The average seeing during the observation was $\sim$0.7 arcsec, and the average humidity fraction $\sim$55.7\%.
The moon illumination was $\sim$16\% and at an angular distance greater than $170\degr$ from the target.

\section{Spectral Analysis} \label{sec:spectal_analysis}
The spectral analysis was performed using the X-ray spectral ﬁtting package \texttt{XSPEC} \citep{Arnaud_1996_XSPEC} version 12.13.1. We applied the interstellar medium abundance and the cross-section tables from \citet{Wilms_2000ApJ} and \citet{Verner_1996ApJ}, respectively. Uncertainties in the spectral parameters are given at 90-percent confidence intervals unless otherwise stated.

\subsection{Broad-band spectral modelling} \label{Sec:broad-band_spectrum}
We first modeled the NuSTAR spectrum of IGR J17498$-$2921 (ObsID 90901317002) in the 3.0–79.0 keV energy band. We included a renormalization factor in the fitting model to account for cross-calibration discrepancies between the two focal plane modules (FPMA and FPMB), ensuring that the values remained consistent within 10\%.
We fixed the absorption column density to the Chandra estimate obtained during the 2011 outburst, $\mathrm{n_H} = (2.87 \pm 0.04) \times 10^{22} \, \mathrm{cm^{-2}}$ \citep{Torres_2011ATel}. 
We first attempted to model the spectra with an absorbed power law but got a poor fit (its reduced chi-squared $\chi^2_r$ being equal to $3.4$ for 1745 degrees of freedom, d.o.f. in the following).
Similarly, an absorbed Comptonized emission model (\texttt{nthComp}; \citealt{Zdziarski_1996MNRAS, Zycki_1999MNRAS}) was tested but alone failed to adequately represent the data ($\chi^2_r=1.4$ for 1743 d.o.f.). 
The residuals indeed suggested the presence of a reflection component. We first added a \texttt{Gaussian} line to fit the broad residuals at the energy of the iron K$\alpha$ emission. The fit $\chi^2$ decreased by $\Delta\chi^2=880$ compared to the previous case, implying a negligible probability that the improvement was due to chance.
The resulting line energy ($(5.8 \pm 0.2)$ keV) was significantly redshifted with respect to the expected rest-frame energy range and had a relatively large width ($(1.8 \pm 0.1)$ keV), suggesting the need for taking relativistic effects into account. The broadness of the line prompted us to attempt modeling it with a \texttt{diskline} component \citep{Fabian_1989MNRAS}. However, the fit always returned extreme values for the parameters controlling the line width, i.e. an inner disk radius of $6 \, \mathrm{R_g}$ (where $\mathrm{R_g}$ is the gravitational radius) and a system inclination of $90^{\circ}$.
We thus adopted a self-consistent reflection model using the \texttt{rfxconv} convolution model \citep{Kolehmainen_2011MNRAS} with the relativistic blurring kernel \texttt{rdblur} \citep{Fabian_1989MNRAS}, integrated with a Comptonized continuum, with parameters fixed to match those of the existing \texttt{nthComp} component within the model \citep[as detailed in previous work; see, e.g.,][]{Anitra_2021A&A}. The reflection fraction is set to negative values so that the \texttt{rfxconv} model returns only the reflection component.
The adopted model was structured as \texttt{constant*TBabs*(nthComp+rdblur*rfxconv*nthComp)}. As recommended in the \texttt {XSPEC} manual, we extended the energy range over which the \texttt{rfxconv} model is calculated up to 500 keV and limited the output spectrum to 79 keV\footnote{\url{https://heasarc.gsfc.nasa.gov/xanadu/xspec/XspecManual.pdf}}.

The Comptonized continuum was described by a photon index of $\Gamma = 1.96_{-0.02}^{+0.01}$, an electron temperature of $\mathrm{kT_e}=17.1_{-0.7}^{+0.9} \, \mathrm{keV}$, a seed photons temperature of $\mathrm{kT_{seed}} = (0.58 \pm 0.02) \, \mathrm{keV}$, and a \texttt{nthComp} normalization of $\mathrm{Norm_{nthComp}}=0.038_{-0.005}^{+0.003}$.
From the \texttt{rdblur} convolution model, we obtained that the index of the power law describing the emissivity was $\beta=-3_{-7}^{+1}$ (with the lower uncertainty indicating that the parameter reached its hard limit of -10), and the inner accretion disk radius was $\mathrm{R_{in}}=13^{+14}_{-7} \, \mathrm{R_g}$ (with the lower uncertainty indicating that the parameter reached its hard limit set at 6 gravitational radii).
We fixed the outer disk radius at 1000 $\mathrm{R_g}$ due to the energy resolution constraints of NuSTAR, as discussed in \citet{DiSalvo_2019MNRAS}, and the system's inclination to $70^{\circ}$ since otherwise it was not well constrained. In the absence of precise inclination estimates for the system (see Sect. \ref{sec:reflection_discussion} for a discussion), we chose $70^{\circ}$ as an upper limit due to the lack of observed eclipses. 
Finally, from the \texttt{rfxconv} model, we obtained a reflection fraction of $-0.5^{+0.1}_{-0.2}$ and an ionization parameter of the disk of $\log{\xi}=3.30^{+0.05}_{-0.09}$, keeping the iron abundance fixed at the solar value.
The $\chi^2$ of the fit was 1732.8 for 1739 d.o.f..

\begin{figure}
   \centering
   \includegraphics[width=0.47\textwidth]{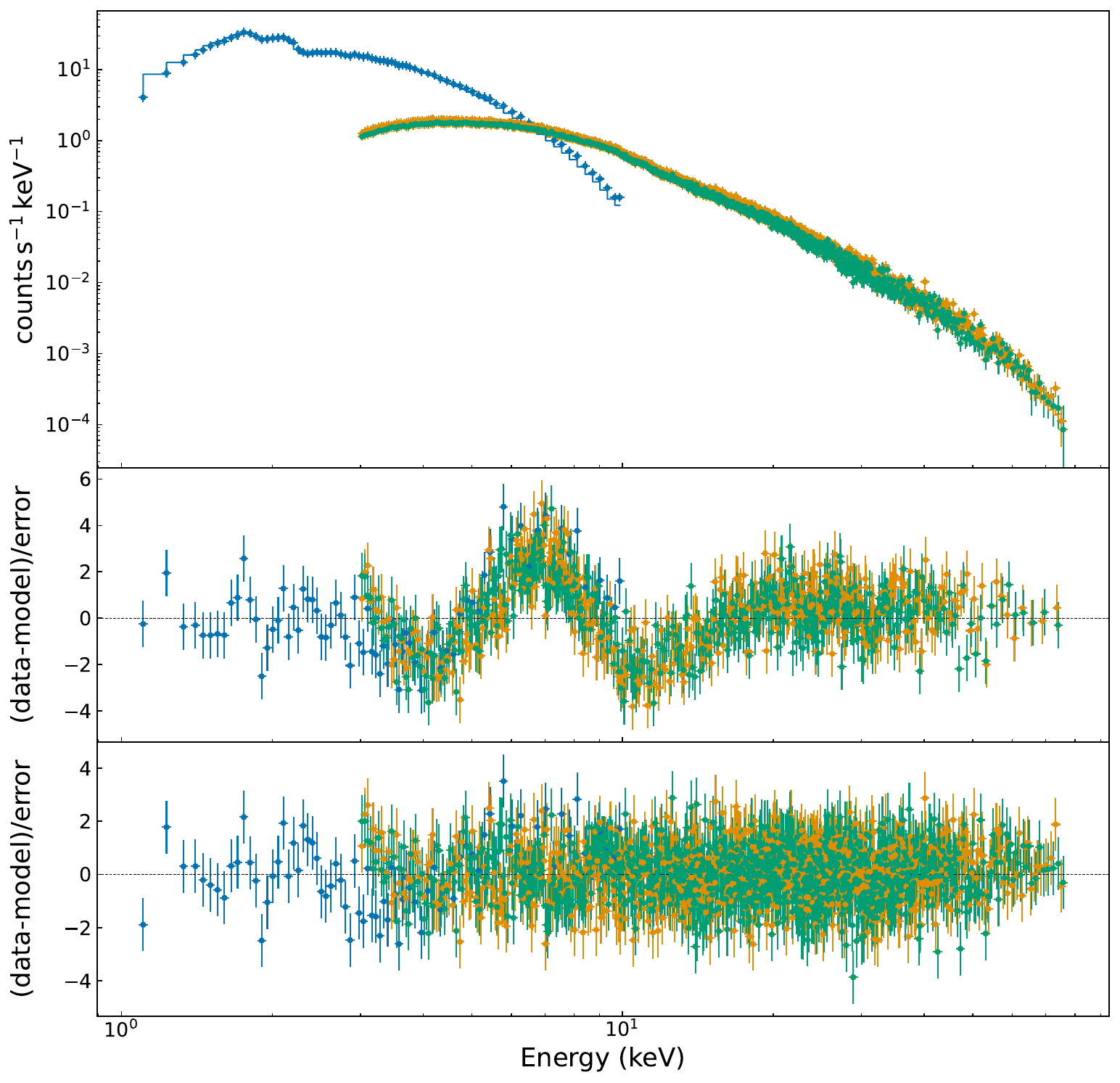}
      \caption{Broad-band X-ray spectrum of IGR J17498$-$2921 as observed on 2023 April 23 by NICER (blue), NuSTAR/FPMA (orange), and NuSTAR/FPMB (green). 
      For NICER we analyzed ObsID 6560010103 in the 1-10 keV energy band, while NuSTAR ObsID 90901317002 is in the 3-79 keV band. The fitted model is  \texttt{constant*TBabs*(gaussian+nthComp+rdblur*rfxconv*nthComp)} (Table \ref{tab:params_broad_band_spectrum}; see Sect. \ref{Sec:broad-band_spectrum} for details). The bottom panel displays the residuals from the fit, while the middle panel shows the residuals obtained after removing the reflection component and repeating the fitting procedure. The data are rebinned for visual purposes.} \label{Fig:total_spectrum}
\end{figure}

Based on the results obtained from the spectral analysis of NuSTAR data, we further investigated the broad-band spectral properties of the source also using quasi-simultaneous observations with NICER (ObsID 6560010103) in the 1.0-10.0~keV energy range.
We did not consider exactly simultaneous NICER observations (ObsIDs 6203770103, 6203770104, 6560010104) because they were affected by a high background count rate, making it unfeasible to constrain the spectral parameters as discussed in Sect. \ref{sec:spectral_evolution}.
To address cross-calibration uncertainties between the two X-ray telescopes, we included a renormalization factor in the model.
We consistently verified that the discrepancies in the values for the constant term obtained from different instruments did not exceed $10\%$.
As mentioned before, Chandra observations of IGR J17498$-$2921 during the 2011 outburst provided an estimate of the absorption column density of $\mathrm{n_H} = (2.87 \pm 0.04) \times 10^{22} \, \mathrm{cm^{-2}}$ \citep{Torres_2011ATel}, later used by \citet{Papitto_2011A&A} for fitting Swift/XRT and RXTE/PCA spectra. On the other hand, \citet{Falanga_2012A&A} studied the broad-band spectra of the source during the same outburst using INTEGRAL/ISGRI, RXTE/PCA, and Swift/XRT data and found smaller values, i.e., $\mathrm{n_H} \sim 1.2-1.5 \times 10^{22} \, \mathrm{cm^{-2}}$. Consequently, given the wide energies range analyzed, we decided to let the neutral absorption column density be free to vary (see Table \ref{tab:params_broad_band_spectrum}).
We added a \texttt{Gaussian} line to every model attempted to account for the residuals at $\sim$$1.6-1.7$ keV in the NICER spectrum (for a discussion on this feature see Sect. \ref{sec:spectral_evolution}). 
To best describe the reflection component observed when fitting NuSTAR spectrum, we adopted again the \texttt{rfxconv} convolution model with the relativistic blurring kernel \texttt{rdblur}. We convolved the reflection component with the \texttt{nthComp} component already present in the model, leading to: 
\texttt{constant*TBabs*(gaussian+nthComp+rdblur*rfxconv*}\\
\texttt{*nthComp)} (see Table \ref{tab:params_broad_band_spectrum} and Fig. \ref{Fig:total_spectrum}).
The best-fit parameters were consistent with those obtained when fitting the model solely to the NuSTAR spectrum. 
Several studies reporting reflection spectra from AMSPs suggested a moderate excess of iron at approximately twice the solar value \citep[see, e.g.,][]{Papitto_2013MNRAS, DiSalvo_2019MNRAS, Sharma_2020MNRAS}.
We tentatively allowed the iron abundance to vary or set it at twice the solar value. However, these attempts did not improve the $\chi^2$ of the fit.
In Table \ref{tab:params_broad_band_spectrum}, the relative fluxes for the Comptonization and the reflection components are calculated in the 1-79 keV range.

\begin{table*} 
\renewcommand{\arraystretch}{1.2}
\centering
\caption{Best-fit parameters for the broad-band spectral model.} \label{tab:params_broad_band_spectrum}
  \begin{tabular}{l c c}          
    \hline\hline
    \multicolumn{3}{c}{\large \textbf{\textsc{Model:}} }
    \rule{0pt}{2.6ex}\\
    \multicolumn{3}{c}{\texttt{constant*TBabs*(gaussian+nthComp+rdblur*rfxconv*nthComp)}}  
    \rule[-1.2ex]{0pt}{0pt} \\
    \hline
    Component & Parameter & Value \\
    \hline
    \scshape{Tbabs} & $\mathrm{n_H}$ ($10^{22}$ cm$^{-2}$) & $2.33^{+0.07}_{-0.05}$\\
    \hline
    \scshape{Gaussian} & $\mathrm{E_{Line}} \, \mathrm{(keV)}$&  $1.61^{+0.04}_{-0.07}$ \\
    & $\sigma \, \mathrm{(keV)}$ & $0.20^{+0.07}_{-0.05}$  \\
    & $\mathrm{Norm_{Line}}$  &$0.004^{+0.002}_{-0.001}$ \\
    \hline
    \scshape{nthComp} & $\Gamma$ & $1.95^{+0.02}_{-0.03}$ \\
    & $\mathrm{kT_e} \, \mathrm{(keV)}$ & $16.5^{+0.9}_{-0.8}$ \\
    & $\mathrm{kT_{seed}} \, \mathrm{(keV)}$ & $0.670^{+0.02}_{-0.01}$ \\
    & $\mathrm{Norm_{nthComp}}$ &$0.026^{+0.002}_{-0.004}$ \\
    & $F_\mathrm{{1-79}} \, \mathrm{(10^{-10} \, erg \, cm^{-2} \, s^{-1})}$ & $6.50 \pm 0.02$ \\
    \hline
    \scshape{reflection model}  &$\mathrm{\beta}$ & $-4^{+2}_{-6^{(\star)}}$  \\
    & $\mathrm{R_{in}}$ ($\mathrm{R_g}$) & $21^{+17}_{-12}$\\
    & $\mathrm{R_{out}}$ ($\mathrm{R_g}$) & $1000^{(*)}$ \\
    & Incl. (degrees) & $\geq 59.8$\\
    & Refl. frac. &  $-0.6^{+0.1}_{-0.2}$ \\
    & Redshift &  $0^{(*)}$ \\
    & Fe abund. (solar units) & $1^{(*)}$ \\
    & $\log{\xi}$ & $3.31^{+0.05}_{-0.08}$ \\
     & $F_\mathrm{{1-79}} \, \mathrm{(10^{-10} \, erg \, cm^{-2} \, s^{-1})}$ & $2.64 \pm 0.02$ \\
    \hline
    \scshape{Total} & $\mathrm{F_{1-79}} \, \mathrm{(10^{-10} \, erg \, cm^{-2} \, s^{-1})}$ & $9.14 \pm 0.02$\\
    \hline
    &$\chi ^2$/d.o.f & 1919.9/1858\\
    \hline
\end{tabular}
\tablecomments{$^{(*)}$ Kept frozen during the fit. 
$^{(\star)}$ The parameter reached its hard limit set at -10.
}
\end{table*}
\noindent

\subsection{Spectral evolution} \label{sec:spectral_evolution}
To search for any variability in the spectral properties of the source during the outburst, we exploited the high-cadence monitoring performed with NICER. We analyzed data acquired from 2023 April 20 (MJD 60054) to 2023 May 3 (MJD 60067; ObsID 6560010113; see Table \ref{table:NICER_obs}). Subsequent observations started from 2023 May 15 (ObsID 6560010114) when the count rate had decreased by a factor greater than about 20 compared to previous observations and the background counts constituted $\sim$70 \% of the total. Owing to the poor statistics, we excluded these observations from the study of the spectral evolution. In our analysis, we paid particular attention to the emission properties in correspondence with the jump of the pulse phase occurring between $\sim$$60060.88$ and $60061.91$ MJD (see Sect. \ref{sec:specral-timing_evolution}).

\begin{figure}
   \centering
   \includegraphics[width=0.45\textwidth]{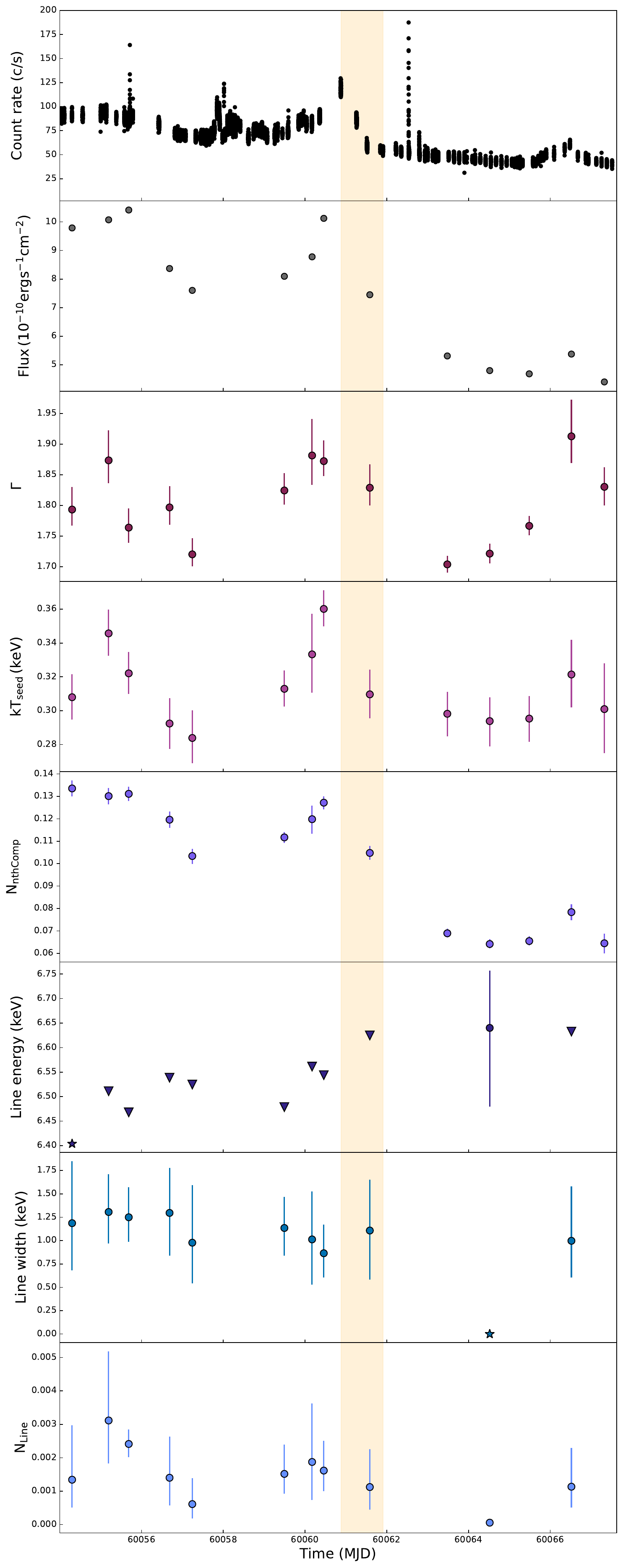}
      \caption{Temporal evolution of the main spectral parameters describing the continuum emission of IGR J17498$-$2921 observed with NICER, along with the iron line. The top panel shows the 0.5-10 keV light curve, and the second panel reports the unabsorbed 1-10 keV flux. 
      The associated uncertainties are reported at a level of statistical confidence of 90 percent. Downward-pointing triangles represent 90 percent c.l. upper limits estimated when a parameter could not be properly constrained, while the star-shaped points represent fixed values for the parameters.
      The spectral parameters are reported in Tables  \ref{Table:spectra_NICER_1} and \ref{Table:spectra_NICER_lines}.
      The yellow-shaded region indicates the time interval in which the phase jump occurs (see Sect. \ref{sec:specral-timing_evolution}).}\label{Fig:main_spectral_evolution}
\end{figure}

We analyzed NICER spectra in the 1.0–10.0 keV band to avoid dealing with instrumental noise present at lower energies \citep[e.g.,][]{Manca_2023MNRAS_LMXB}.
Observations with ObsIDs 6203770103, 6203770104, 6560010104, and 6560010108 were affected by a high background. While in the other NICER observations the background counts constituted about 1-2\% of the total, in these particular observations the background counts increased to $\sim$9\%, 10\%, 14\%, and 7\%, respectively.
Attempts to fit these spectra, even by fixing some parameters based on previous estimates, resulted in a systematically lower photon index ($\Gamma \sim 1.5$, whereas the weighted average obtained from the other spectra is ($1.765 \pm 0.009$); see Table \ref{Table:spectra_NICER_1} and Fig. \ref{Fig:main_spectral_evolution}). To ensure consistency in studying spectral variability during the outburst, we opted to exclude these observations from our analysis.

The continuum was well described by an absorbed Comptonized emission.
Due to NICER lack of spectral coverage above 10 keV, we were unable to accurately determine the electronic temperature. Therefore, we set $\mathrm{kT_e}$ to 20 keV for all spectra, following the result obtained from the broad-band spectral analysis (Table \ref{tab:params_broad_band_spectrum}) and aligned with the typical values observed in similar sources.
We fix the neutral absorption column density at the Chandra estimate of $\mathrm{n_H} = (2.87 \pm 0.04) \times 10^{22} \, \mathrm{cm^{-2}}$ \citep{Torres_2011ATel} in order to better constrain the other spectral parameters and investigate a possible spectral evolution.

\begin{figure}
   \centering
   \includegraphics[width=0.47\textwidth]{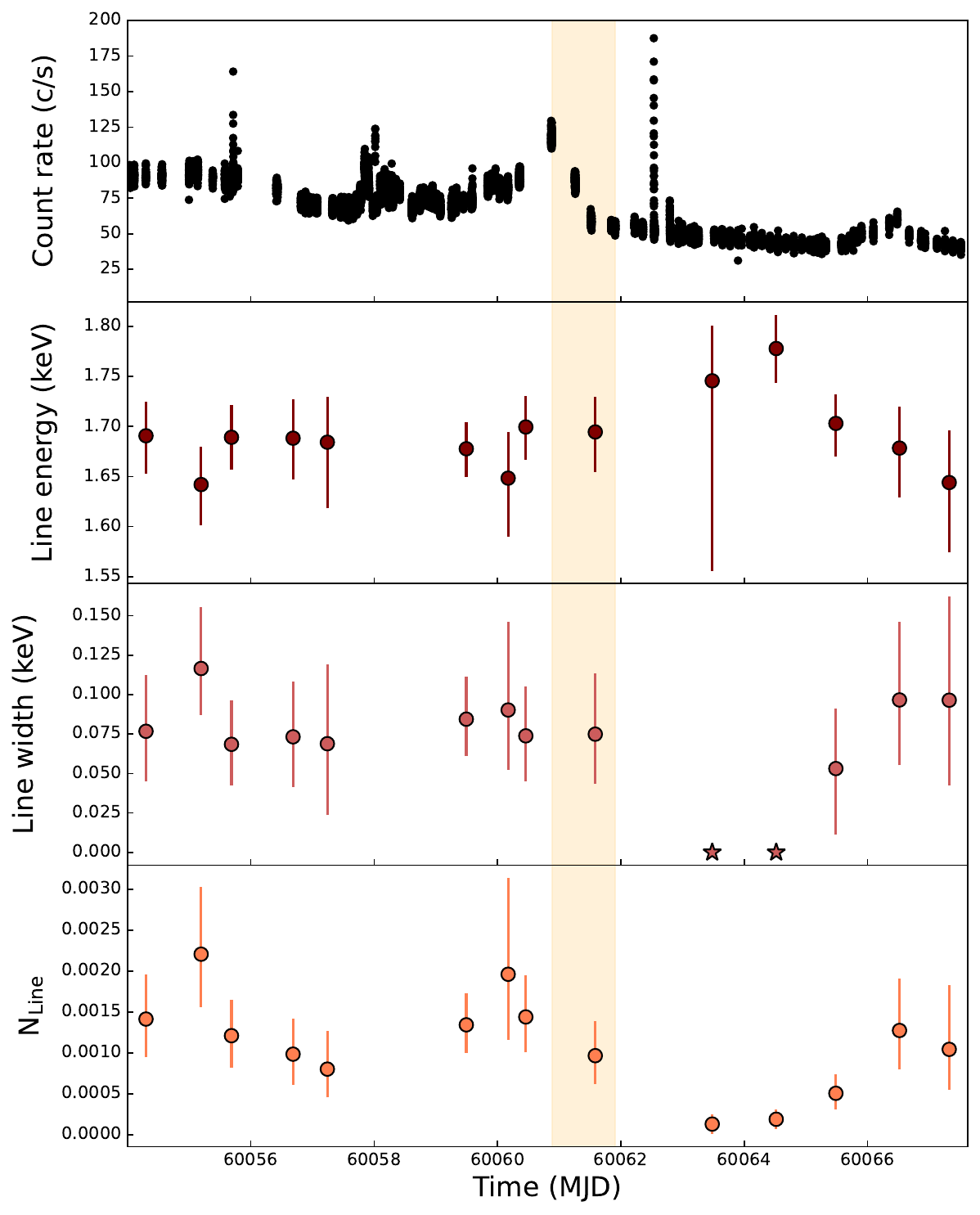}
      \caption{Temporal evolution of the likely Si fluorescence line at $\sim$1.7 keV.
      The top panel shows the 0.5-10 keV light curve.
      The associated uncertainties are reported at a level of statistical confidence of 90 percent. Star-shaped points represent fixed values for the parameters. 
      The spectral parameters are reported in Table \ref{Table:spectra_NICER_lines}.
      The yellow-shaded region indicates the time interval in which the phase jump occurs (see Sect. \ref{sec:specral-timing_evolution}).}\label{Fig:gaussian_Si_line}
\end{figure}

The inclusion of a \texttt{Gaussian} line centered at $\sim$1.7 keV (see Fig. \ref{Fig:gaussian_Si_line}) always significantly improved the data modeling (F-test probability always lower than $\sim$$9 \times 10^{-4}$; while typically not appropriate for confirming the addition of a Gaussian component, as outlined by \citealt{Protassov_2002ApJ}, extending the normalization range to negative values permits the application of the F-test statistic to demonstrate improved fit, as noted by, e.g., \citealt{Anitra_2021A&A}).
Modeling NICER spectra of the neutron star low-mass X-ray binary SWIFT J1749.4$-$2807, \citet{Marino_2022MNRAS} added a Gaussian line at $\sim$1.7 keV, reporting that this feature is most likely a Si fluorescence line from the Focal Plane Modules (according to private communication with M. Corcoran).

We also tried to add a black body to our model, but it was undetectable likely due to the limited statistics of the NICER spectra. Despite the F-test indicating that including this component significantly improved the $\chi^2$ of most fits, the resulting best-fit parameters deviated considerably from those typically observed in other AMSPs. Specifically, the black-body temperature reached up to 2 keV, while the normalization was very low, corresponding to a radius of $\sim$1 km (assuming a distance of 7.6 kpc) or less. This suggested that the black-body component was not addressing a soft excess but rather fitting residuals of the Comptonization component. Consequently, we decided not to include this component.

Given the prominent reflection component observed in the broad-band spectrum (Sect. \ref{Sec:broad-band_spectrum}) and the spectral stability shown by this source, we attempted to account for any residual features in the energy range where the iron line is expected, i.e. $\sim$6.40-6.97 keV, with a \texttt{Gaussian} line. 
Following the procedure described in recent works \citep[e.g.,][]{Manca_2023MNRAS_AMXP}, where the significance of the iron line is assessed by computing the ratio between its normalization and the associated uncertainty at $1\sigma$ confidence level, we found that in all spectra the normalization did not deviate from the continuum by more than $\sim$$3\sigma$. However, we verified that the addition of this \texttt{Gaussian} line did not significantly change the other parameters, so we added it to the model owing to the strong reflection observed in the broad-band spectrum (see Fig. \ref{Fig:main_spectral_evolution}).

In summary, the model that best describes the majority of NICER spectra was \texttt{TBabs*(nthComp+gaussian+gaussian)}. The best-fit spectral parameters are reported in Tables \ref{Table:spectra_NICER_1} and \ref{Table:spectra_NICER_lines}.

\subsection{Correlated spectral--timing evolution} \label{sec:specral-timing_evolution}
To study the correlated evolution of the spectral and timing properties of IGR J17498$-$2921 during its 2023 outburst, we provide a preliminary timing analysis of the coherent pulsations based on NICER data. We refer to 
\citet{Strohmayer_in_prep} for a detailed analysis of the timing properties of the source during this outburst.

\begin{figure}
   \centering
   \includegraphics[width=0.47\textwidth]{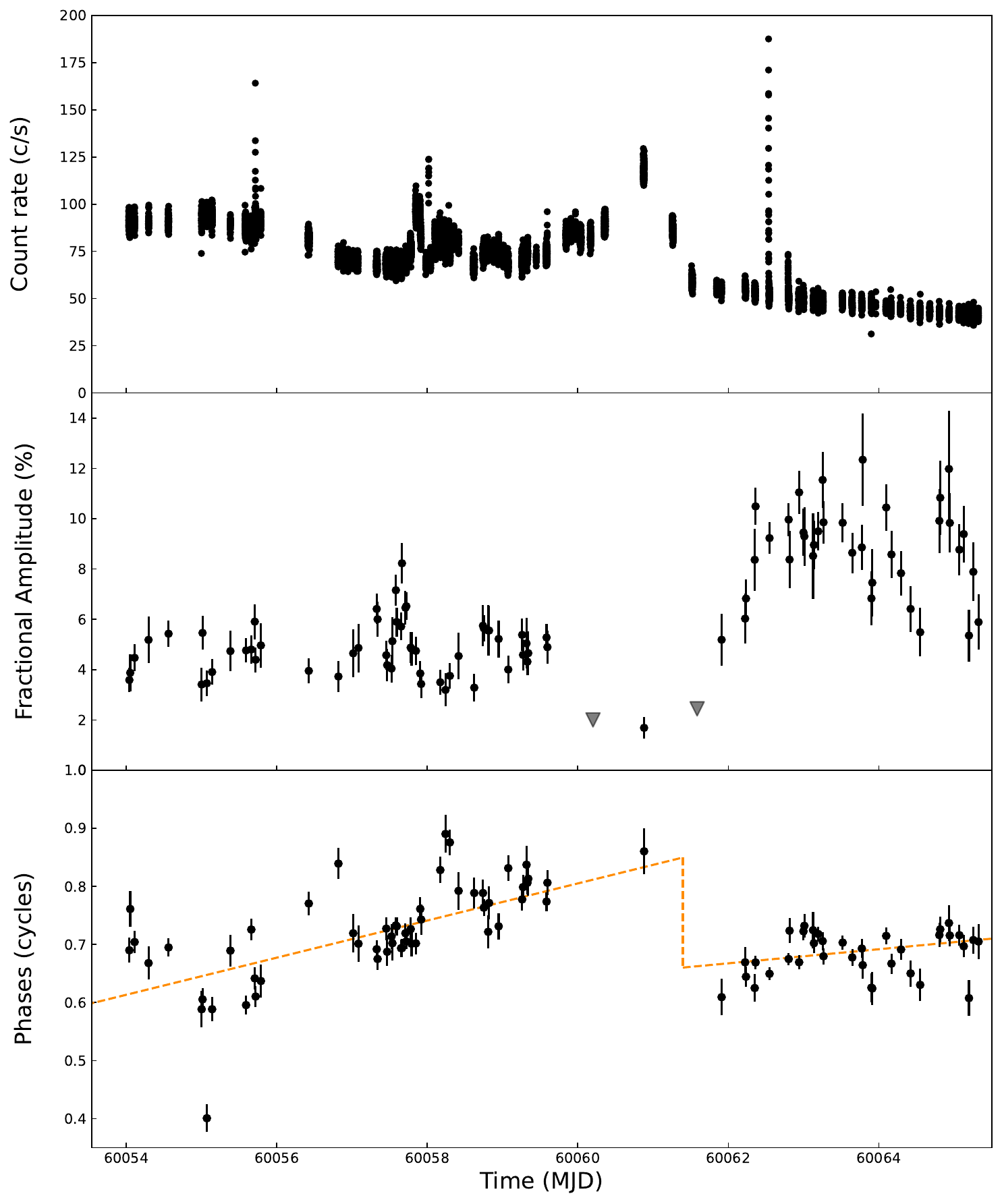}
      \caption{Temporal evolution of the 2023 outburst monitored with \textrm{NICER}. \textit{Top panel}: the 0.5--10~keV light curve using 10-s bins. The type-I X-ray burst (Fig. \ref{Fig:lc_and_bursts}) was filtered out to perform the timing analysis. \textit{Second panel}: pulse fractional amplitude. \textit{Bottom panel}: pulse phase. The orange dotted line represents a step function with a linear fit applied to the data before and after the phase jump.} \label{Fig:timing_ampl_phases}
\end{figure}

In order to correct the photon arrival times for the pulsar orbital motion in the binary system, we first propagated the orbital parameters found in \citet{Papitto_2011A&A}. We divided our data into 1000-s long segments and folded them around our best estimate of the pulsar spin frequency, using 16 phase bins. We modeled the pulse profiles with a constant plus the fundamental harmonic term, retaining only the data in which the signal was detected with an amplitude significant at more than a 3$\sigma$ confidence level. Table \ref{table:Timing_sol} shows the best-fitting orbital and spin parameters we obtained by modeling the time evolution of the pulse phase computed over the first harmonic with a constant frequency solution, $\bar{\nu}$, \citep[see, e.g.,][ and references therein]{Illiano_2023ApJ}. We estimated the systematic uncertainty on the spin frequency due to the positional uncertainties of the source \citep[see, e.g.,][]{Lyne_GrahamSmith_1990, Burderi_2007, Sanna_2017}. Adopting the positional uncertainties of 0.4$^{\prime\prime}$ at 1$\sigma$ confidence level reported by \citet{Papitto_2011A&A}, we estimated $\sigma_{\nu_{\textrm{pos}}} \leqslant 7.8 \times 10^{-8} \, \mathrm{Hz}$. 
\begin{table} 
\renewcommand{\arraystretch}{1.2}
\centering
\caption{Preliminary timing solution for IGR J17498$-$2921 during the 2023 outburst.} 
\label{table:Timing_sol}      
\begin{tabular}{l c}          
\hline\hline                       
Parameter & Value\\
\hline
Epoch (MJD) & 60054.0349452\\
$P_{\textrm{orb}}$ (s) & 13835.616(70)\\
$a_1 \sin{i}$ (lt-s) & 0.365201(23)\\ 
$T_{\textrm{asc}}$ (MJD) & 60054.0855811(37)\\
\hline
$\bar{\nu}$ (Hz) & 400.990186043(23)\\
$\bar{\chi} ^2$/d.o.f. & 916.8/88\\
\hline
$\nu_{1}$ (Hz) & 400.990185623(58)\\
$\chi_{1} ^2$/d.o.f. & 372.1/52\\
\hline
$\nu_{2}$ (Hz) & 400.990185910(62)\\
$\chi_{2} ^2$/d.o.f. & 80.3/31\\
\hline
\end{tabular}
\tablecomments{The timing solution was obtained adopting the source coordinates from \citet{Chakrabarty_2011ATel}.
$P_{\textrm{orb}}$ is the orbital period, $a_1 \sin{i}$ is the projected semimajor axis, and $T_{\textrm{asc}}$ is the epoch of passage at the ascending node.
To take into account the large value of the reduced $\chi^2$ obtained from the fit, we rescaled the uncertainties of the fit parameters by the square root of that value \citep[see, e.g.,][]{Finger_1999}. Uncertainties are the 1$\sigma$ statistical errors.}
\end{table}
\noindent
Our results are compatible with those of \citet{Strohmayer_in_prep}. 
We observed a phase jump of $\sim$-0.25 cycles in the phases occurring between 60060.881021--60061.911115 MJD (bottom panel of Fig. \ref{Fig:timing_ampl_phases}), just after an episode of increased flux. We also found that in correspondence with the X-ray flare the amplitude of the pulsations dropped below the detection threshold (middle panel of Fig. \ref{Fig:timing_ampl_phases}, where the triangles represent the $3\sigma$ upper limits computed following the procedure in \citealt{Vaughan_1994}). We modeled data separately taken before and after such phase jump. The two solutions, with the two values of the spin frequency indicated in Table \ref{table:Timing_sol} as $\nu_1$ and $\nu_2$, respectively, have a reduced $\chi^2$ lower than that of the overall solution.

During the 2011 outburst of IGR J17498$-$2921, \citet{Papitto_2011A&A} estimated a spin frequency of $400.99018734(1)$ Hz and a first derivative equal to $-6.3(1.1) \times 10^{-14}$ Hz/s at the reference epoch 55786.124 MJD. Assuming that the NS kept spinning down at this rate during the outburst until it returned to quiescence on 2011 September 19 (MJD 55823; \citealt{Linares_2011ATel_quiescence}), the spin frequency should have varied to $400.99018714(4)$ Hz. Using the timing solution obtained before the phase jump (i.e., the spin frequency $\nu_1$ in Table \ref{table:Timing_sol}), we obtained a long-term spin derivative of $\dot{\nu}_{\mathrm{SD}} = -4.1(2) \times 10^{-15} \, \mathrm{Hz/s}$. This value is of the same order of magnitude as the secular spin-down observed in other AMSPs, such as SAX J1808.4$-$3658 \citep{Illiano_2023ApJ}, IGR J00291$+$5934 \citep{Papitto_2011A&A_IGRJ00291}, and XTE J1751$-$305 \citep{Riggio_2011A&A}, and compatible with the expected emission from a $\sim$10$^8$ G rotating magnetic dipole (see Sect. \ref{sec:phase_jump_discussion} for a detailed discussion).

We investigated potential spectral changes around the peculiar phase jump of $\sim$-0.25 cycles around 60060.881021--60061.911115 MJD. Fitting the time evolution of the main spectral parameters with a constant model revealed variations in the photon index, $\Gamma$ ($\chi^2_r \sim 6.0$ for 13 d.o.f.), and the seed photons' temperature, $\mathrm{kT_{seed}}$ ($\chi^2_r \sim 2.8$ for 13 d.o.f.). Specifically, both the $\Gamma$ and $\mathrm{kT_{seed}}$ tracked the flux variations showed during and after the X-ray flare that preceded the phase jump (see Fig. \ref{Fig:main_spectral_evolution}). 

\section{Fast optical variability}
\label{sec:optical_pulsations}
Motivated by the discovery of optical pulsations from a transitional \citep{Ambrosino17} and an accreting millisecond pulsar \citep{Ambrosino21}, we searched for a coherent signal in the Leahy-normalized Fourier power spectrum \citep{Leahy_1983ApJ} of SiFAP2 data obtained after correcting the photons' times of arrival with the orbital parameters listed in Table~\ref{table:Timing_sol}. Fitting the power spectrum in the range 450--800~Hz (i.e., not including the frequencies where the harmonics of the coherent signal are expected) with a constant, we realized that both the average and standard deviation of the noise level were $r \simeq 2.26$, significantly higher than the expected value of $\bar{r}=2$.  The origin of such a discrepancy was likely related to a higher noise variance due to detector cross-talk and will be addressed in a separate paper (La Placa et al., in prep.). Here, we rescaled the power observed in each frequency bin by the ratio $r/\bar{r}$ to avoid 
overestimating the significance of any putative signal.
The frequency resolution of a power spectrum performed on the whole observation length ($t_{\mathrm{obs}}=4.5$~ks, $\delta\nu=2\times10^{-4}$~Hz) is much coarser than the accuracy in the available pulsar ephemeris (see Table~\ref{table:Timing_sol}). Therefore, the search for a coherent signal in the optical band at the known spin frequency of the NS was limited to a single-trial search. As a result, the power threshold for a signal to have a probability of less than $p=2.7 \times 10^{-3}$ of being due to chance\footnote{The false alarm probability, denoted by $p$, is defined as $1-C$, where $C$ is confidence level set at 3$\sigma$ \citep[see, e.g.,][]{Vaughan_1994}.} was $\sim$$2 \ln{(1/p)} \simeq 11.8$. Given the total number of counts observed ($\sim$$1.3\times10^8$), the power value of 3.3 we observed translated into an upper limit of $\sim$0.06\% on the pulse amplitude (evaluated according to \citealt{Vaughan_1994}).
The upper limit on the background-subtracted amplitude was likely much higher. In fact, the background observed from a region located 25 arcsec away from the target (i.e., outside the $7 \times 7$~arcsec region observed when pointing at the source position) towards the east direction for 150 s ($\sim$$2.91 \times 10^{4}$ c/s) was compatible with the count rate observed from the source.
During the 2011 outburst, an optical counterpart with a magnitude of $\mathrm{i'} = 22.5 \pm 0.7$ was proposed \citep{Russell_2011ATel}. Considering the Galactic hydrogen column density observed towards the source ($\mathrm{n_H}=2.87(4)\times10^{22}$~cm$^{-2}$), we expected an extinction of $\simeq$10 and 5 mag in the V and i$^{'}$ bands, respectively \citep{cardelli89, Foight_2016ApJ}. 
Assuming the same optical spectrum of PSR J1023+0038 \citep{cotizelati14}, we then expected a low observed V magnitude ($V_{\mathrm{src}}\simeq27.5$ mag) 
due to interstellar absorption. The expected SiFAP2 source countrate\footnote{\url{https://www.tng.iac.es/instruments/sifap2/}} ($\sim$1 c/s) was thus much lower than the background and prevented us from finding a meaningful upper limit on the background-subtracted pulse amplitude.

\section{Discussion and conclusions} \label{sec:discussion}

\subsection{Phase jump} \label{sec:phase_jump_discussion}
The coherent timing analysis of X-ray pulsations from IGR J17498$-$2921 during its 2023 outburst highlighted the occurrence of a phase jump of $\sim$-0.25 cycles just after a X-ray flare coinciding with a decrease in the pulse amplitude below the $3\sigma$ significance level (see Fig. \ref{Fig:timing_ampl_phases} and \citealt{Strohmayer_in_prep}). 
Erratic variations in the X-ray phases of AMSPs during outbursts, commonly referred to as timing noise \citep[see][for a review and references therein]{Patruno_Watts_2021ASSL}, have been documented in different sources, such as SAX J1808.4$-$3658 \citep{Burderi_2006ApJ, Hartman_2008ApJ}, XTE J1814$-$338 \citep{Papitto_2007MNRAS}, XTE J1807$-$294 \citep{Riggio_2008Ap, Chou_2008ApJ}, and MAXI J1957$+$032 \citep{Sanna_2022MNRAS_MAXIJ1957}. The origins of timing noise are not yet fully understood, but potential causes include instabilities in the accretion flow \citep[e.g.,][]{Burderi_2006ApJ, Papitto_2007MNRAS, Patruno_2009ApJ}, disk-magnetosphere interactions \citep{Kajava_2011MNAS}, and changes in the neutron star's internal dynamo processes \citep{Long_2012NewA}. These irregularities hinder the understanding of spin evolution, making it challenging to disentangle contributions from accretion torques and other stochastic processes.

Particular attention has been paid to the hypothesis that phase jumps could stem from hot spot drifts on the NS surface, as reproduced in magnetohydrodynamic (MHD) simulations \citep[e.g.,][]{Romanova_2004ApJ, Kulkarni_2009MNRAS, Kulkarni_2013MNRAS}. \citet{Lamb_2009ApJ} proposed that, when the NS magnetic and rotational axes are almost aligned, slight variations in the mass accretion rate result in inflowing matter being channeled along significantly different sets of field lines, producing comparatively large displacements of the emitting regions on the NS surface. 
According to the ``nearly aligned moving spot'' model, an increase in pulse amplitude should correspond to a decrease in phase scatter. This phenomenon can be understood by considering two main factors. Firstly, when a spot is centered on the spin axis, its emission is axisymmetric, leading to no modulations. As the emitting spot gets farther from the spin axis, it becomes slightly asymmetric, producing weak modulations. Consequently, even a small shift in latitude can greatly affect the pulse amplitude. Secondly, a small displacement in the azimuthal position of the emitting spot causes a significantly larger phase shift when it is near the spin axis compared to when it is farther away. 
Our observations align with this expectation. 
As shown in Fig. \ref{Fig:timing_ampl_phases}, the pulse amplitude ranges from $\sim$3--6\% before the phase jump ($\sim$60054--60061 MJD) to $\sim$5--12\% after it (after $\sim$60062 MJD). This variation in coincidence with a decrease in X-ray flux has already been observed in several AMSPs, such as MAXI J1957$+$032 \citep{Sanna_2022MNRAS_MAXIJ1957}, SAX J1808.4$-$3658 \citep{Bult_2022ApJ, Illiano_2023ApJ}, and MAXI J1816$-$195 \citep{Bult_2022ApJ}. In our case, before the phase jump, there is a noticeable phase scatter when the pulse amplitude is lower, whereas afterward higher amplitudes are accompanied by more stable phases. The root mean square of phase residuals obtained using the corresponding timing solutions before and after the phase jump (see Table \ref{table:Timing_sol}), are $(0.063 \pm 0.009)$ and $(0.036 \pm 0.006)$ cycles, respectively, confirming this transition.
As mentioned before, \citet{Lamb_2009ApJ} suggest that the non-detectability of pulsations can be attributed to a hot spot approaching the spin axis. However, it is not easily understood how this could be correlated with an increase in the mass accretion rate. According to the standard scenario of accreting pulsars, a higher X-ray luminosity traces an increase in the mass accretion rate and a shrinkage of the inner radius of the accretion disk. 
Consequently, in the case of a misalignment between the spin and the magnetic axis, matter would attach to field lines leading to a location on the NS surface farther from the spin axis, rather than closer.
This scenario is inconsistent with our findings shown in Fig. \ref{Fig:timing_ampl_phases}, where the drop in the pulse amplitude coincides with a X-ray flare. Following this episode, the pulse amplitude starts to increase again, and the X-ray flux decreases, coinciding with the phase jump of $\sim$-0.25 cycles. This phase jump and the subsequent increase in pulse amplitude, despite the declining X-ray flux, suggest complex dynamics not fully explained by the current model expectations.

The non-detection of pulsations during the X-ray flare, preceded and followed by the presence of pulsations at lower X-ray luminosity, may be explained by a magnetic field buried by the higher mass accretion rate \citep[e.g.,][]{Cumming_2001ApJ}. 
In Sect. \ref{sec:specral-timing_evolution}, we computed the long-term spin frequency evolution to be $\dot{\nu}_{\mathrm{SD}}=-4.1(2) \times 10^{-15}$ Hz/s. To estimate the NS magnetic field and, consequently, the truncation radius of the disk during the X-ray flare, we assume, as a first-order guess, that Larmor’s formula holds for the NS regarded as a magneto-dipole rotator during quiescence. 
The energy loss rate due to magnetic-dipole radiation in the limit of force-free magneto-hydrodynamics is given by $\dot{E}_{\mathrm{dip}}=-(1+\sin^2{\alpha})\, (B R^3)^2 \, (2 \pi \nu)^4/c^3$ \citep{Spitkovsky_2006ApJ}, where $\alpha$ is the magnetic inclination angle, $B$ is the dipolar NS surface magnetic field, $R$ is the NS radius (assumed to be 10 km), and $c$ is the speed of light.
We equate this expression to the loss rate of rotational kinetic energy, $\dot{E}_{\mathrm{kin}}=-4 \pi^2 I \nu \dot{\nu}$ \citep[e.g.,][]{Spitkovsky_2006ApJ}, where $I$ is the NS moment of inertia (assumed to be $10^{45} \, \mathrm{g \, cm^{2}}$). For $\alpha=0$, this translates into an upper limit for the NS magnetic field of $B \lesssim 2 \times 10^{8} \, \mathrm{G}$.
To compute the mass accretion rate during the X-ray flare, we extrapolated the unabsorbed 0.5-100 keV flux, $F_X \sim 3 \times 10^{-9} \, \mathrm{erg \, cm^{-2} \, s^{-1}}$, and adopted a distance of $\sim$7.6 kpc \citep{Linares_2011ATel}. We obtained $\dot{M} \simeq 1 \times 10^{17} \, \mathrm{g \, s^{-1}}$.
Assuming that the disk is truncated at a radius equal to $\xi$ times the Alfvén radius \citep[e.g.,][]{Ghosh_Lamb_1979ApJ}, $R_{\mathrm{m}} = \xi_{0.5} \, [(B\,R^3)^4/(2\,G\,M\,\dot{M}^2)]^{1/7}$, where $\xi_{0.5}=(\xi/0.5)$, and $G$ is the gravitational constant. For a 1.4 M$_\odot$ NS with a 10 km radius, we have $R_{\mathrm{m}} \sim 13$ km.
The hypothesis that pulsations are not detectable due to an increased mass accretion rate, resulting in a disk close to the NS surface, cannot be excluded. However, we note that in the 0.5-10 keV light curve the count rate varies only slightly, from about $\sim$90-100 c/s at the beginning of the outburst to $\sim$120 c/s during the X-ray flare. This, combined with the estimate of the inner disk radius obtained from modeling the reflection component (see Table \ref{tab:params_broad_band_spectrum}) suggests that these considerations should be approached with caution.

In some AMSPs, flux-phase (anti-)correlations have been observed \citep[e.g.,][]{Patruno_2009ApJ} and recently proposed as a tool to constrain hot spot displacements. \citet{Bult_2020ApJ} described the phase evolution during the 2019 outburst of SAX J1808.4$-$3658 with a flux-dependent timing model, assuming that phase variations are influenced by hot spot drifts related to changes in the inner disk radius as $\phi \propto \mathrm{R_{in}} \propto \dot{M}^{-1/5}$. This model is not consistent with our observations, as the phase jump of {$\sim-$0.25} does not appear to be anti-correlated with the X-ray flux.
It is also noteworthy that during the subsequent 2022 outburst of SAX J1808.4$-$3658, the observed anti-correlation between the X-ray flux and the phases was likely not due to $\dot{M}$-related hot spot drifts \citep{Illiano_2023ApJ}, as was similarly observed for the AMSP MAXI J1816$-$195 \citep{Bult_2022ApJ}. The best-fit exponential indices of the flux-dependent timing model were not consistent with the expectations from the aforementioned MHD simulations of moving hot spots.

A comprehensive explanation for the physical origin of the X-ray phase wandering is still lacking, making it necessary to carry on coherent timing analysis of AMSPs in outbursts to investigate accretion theories. We speculate that the observed phase jump, along with an increase in the pulse amplitude and a decrease in the X-ray flux after the flare, may be linked to a change in the accretion geometry, potentially involving a reconfiguration of the magnetic field lines. This prompted us to investigate potential correlated variability in the timing and spectral properties of the source, as discussed in the next section.

\subsection{X-ray Spectral properties}
Throughout their outbursts, AMSPs commonly display hard X-ray spectra \citep[see, e.g.,][and references therein]{Poutanen_2006AdSpR, DiSalvo_2023hxga.book}, where the photon flux density is typically described by a power law $dN/dE \propto E^{-\Gamma}$ with a photon index $\Gamma$ falling within the range of  $\sim$1.8-2.0. This hard component is usually modeled with a thermal Comptonization, where electrons featuring temperatures of $\sim$20-50 keV up-scatter seed black body-like photons with a temperature of $\sim$0.3 - 1.0 keV \citep[see, e.g.,][]{Papitto_2020NewAR}.

Our spectral analysis using NICER and NuSTAR (Sects. \ref{Sec:broad-band_spectrum} and \ref{sec:spectral_evolution}) indicated that the spectral continuum of IGR J17498$-$2921 during its 2023 outburst aligned well with the typical emission observed in AMSPs. The \texttt{nthComp} model yielded a photon index in the range of $\sim1.7-2.0$ (Tables \ref{tab:params_broad_band_spectrum} and \ref{Table:spectra_NICER_1}).
The estimated electron temperature was found to be $\mathrm{kT_e}\sim17$ keV for both NuSTAR and the broad-band spectral analysis, while it was fixed at 20 keV during NICER spectral fitting due to the lack of spectral coverage beyond $\sim$10 keV.
The seed temperature in the broad-band spectral analysis was $\mathrm{kT_{seed}}\sim0.7$ keV, whereas it appeared lower in the NICER spectra ($\mathrm{kT_{seed}}\sim0.3$ keV), but within the typical values for AMSPs.

In the Comptonization model \texttt{nthComp}, the asymptotic power-law index ($\Gamma$) is related to the temperature of the hot Comptonizing electrons ($\mathrm{kT_e}$) and the medium optical depth $\tau$ through
the relation \citep{Zdziarski_1996MNRAS, Lightman_1987ApJ}
\begin{equation}
    \Gamma=\left[ \frac{9}{4}+\frac{1}{\tau\left(1+\frac{\tau}{3}\right)\left(\frac{\mathrm{kT_e}}{m_ec^2}\right)}\right]^{1/2}-\frac{1}{2}.
\end{equation}
Using the result shown in Table \ref{tab:params_broad_band_spectrum} and the previous expression, we estimated the medium optical depth to be $\tau \sim 3.7$.
This result falls at the upper end of the typical values for the Comptonizing medium in AMSPs ($\tau \sim 1.6-3.7$ based on the previously mentioned common ranges of $\Gamma \sim 1.8-2.0$ and $\mathrm{kT_e} \sim 20-50$ keV). 
A high value for the optical depth of the Comptonization corona is compatible with the low electron temperature reported in Table \ref{tab:params_broad_band_spectrum}, indicating a significant number of scatterings for the incoming photons.

Typically, in the spectra of AMSPs two weak soft thermal components are observed to contribute modestly to the source luminosity, accounting for $\sim$10-20\% of the total flux \citep{DiSalvo_2023hxga.book}. They are generally modeled as thermal emission from a cooler ($\sim$0.4-0.6 keV) accretion disk and a hotter ($\sim$1 keV) spot on the NS surface \citep{Poutanen_2006AdSpR}.
However, in our work, although it is clear that accretion is occurring due to the observed X-ray outbursts, the NICER spectra were predominantly dominated by the Comptonization component, with no thermal emission detected.
Attempts to fit the weak black-body component yielded physically unreliable parameters, indicating that the additional component was contributing to the Comptonization fit. This phenomenon, where the black-body component is too faint to be detected or radiates below the observed energy range, has been noted in other sources \citep{DiSalvo_2023hxga.book}. Therefore, we did not include the black body component in our spectral fits.

In Sect. \ref{sec:specral-timing_evolution}, we investigated potential spectral variability around the phase jump of $\sim$-0.25 cycles. 
As shown in Fig. \ref{Fig:main_spectral_evolution}, we observed changes in both the photon index and the temperature of the seed photons that mirrored variations in the X-ray flux. This evolution was particularly pronounced around the phase jump, where we noted a steepening of the power law and an increase in the seed photon temperature coinciding with the X-ray flare. Subsequently, these parameters decreased.

\subsection{Reflection spectrum} \label{sec:reflection_discussion}
The reflection component plays a crucial role in constraining the system's geometry, such as the inclination and the inner accretion disk radius \citep[e.g.,][]{Fabian_1989MNRAS, Fabian_2010SSRv}.
Reflection spectra from AMSPs have been observed, e.g., from SAX J1808.4$-$3658 \citep{Cackett_2010ApJ, Papitto_2009A&A, Patruno_2009MNRAS, Wilkinson_2011MNRAS, DiSalvo_2019MNRAS}, HETE J1900.1$-$2455 \citep{Cackett_2010ApJ, Papitto_2013MNRAS}, IGR J17511$-$3057 \citep{Papitto_2010MNRAS, Papitto_2016A&A}, SAX J1748.9$-$2021 \citep{Pintore_2016MNRAS, Sharma_2020MNRAS}, IGR J0029$+$5934 \citep{Sanna_2017MNRAS}, and the eclipsing SWIFT J1749.4$-$2807 \citep{Marino_2022MNRAS}. However, not all AMSPs exhibited reflection features \citep[e.g.,][]{Falanga_2005A&A, Sanna_2018A&A}.

We derived a system's inclination $\geq 59.8^{\circ}$ within the \texttt{rfxconv} model. Despite not being well constrained, this value is consistent with the expectation that the system's inclination should be below 80$^{\circ}$ as no eclipses have been observed. 
\citet{Falanga_2012A&A} inferred a low inclination for IGR J17498$-$2921 from burst energetics analysis, emphasizing the relative beaming of persistent and burst emissions. They found the nuclear energy release ($Q_{\mathrm{nuc}}$) to be lower than expected for pure helium fuel, yet the observed short burst rise times and durations indicated hydrogen-poor material at ignition. This discrepancy was attributed to a persistent emission more strongly beamed towards the observer than the burst one, typical of low inclination systems, aligning $Q_{\mathrm{nuc}}$ with expectations for pure helium fuel.
Additionally, \citet{Papitto_2011A&A} estimated a lower limit on the inclination equal to $24.6^{\circ}$ for a NS mass of $1.4 \, \mathrm{M_\odot}$ by assuming that the companion star fills its Roche Lobe and follows the zero-age main sequence mass-radius relation. As the companion stars of AMSPs are generally bloated compared to the thermal equilibrium radius, a larger value of the inclination is likely. While our best-fit inclination deviates from past estimates, it falls within the range indicated by \citet{Papitto_2011A&A}.

The ionization parameter of the disk is given by $\xi = L_X/(n_e \, r^2)$, where $L_X$ is the luminosity of the X-ray incident spectrum, $r$ is the distance between the illuminating source and the emitting region, and $n_e=1.2\, n_H$ with $n_e$ and $n_H$ being the electron and hydrogen number densities in the emitting region, respectively \citep{Garcia_2013}. 
We extrapolated the unabsorbed 0.5-100 keV flux, $F_X = (1.892 \pm 0.004) \times 10^{-9} \, \mathrm{erg \, cm^{-2} \, s^{-1}}$, and assumed a NS radius of 10 km and a distance of $\sim$7.6 kpc \citep{Linares_2011ATel}. We have $L_X \simeq 1.3 \times 10^{37} \, \mathrm{erg \, s^{-1}}$. Using a distance of $r \sim 20 \, \mathrm{R_g}$ and our best-fit ionization parameter of $\log \xi \sim 3.3$ (see Table \ref{tab:params_broad_band_spectrum}), we have $n_e \simeq 4 \times 10^{22} \, \mathrm{cm^{-2}}$ and $n_H \simeq 3.4 \times 10^{22} \, \mathrm{cm^{-2}}$. 

The best-fit value for the inner disk radius is $\mathrm{R_{in}} = 21^{+17}_{-12} \, \mathrm{R_g}$, which corresponds to $\mathrm{R_{in}} = 43^{+35}_{-25}$ km for a 1.4 $\mathrm{M_\odot}$ NS.
Estimations of the inner disk radius are pivotal for investigating accretion theories on rapidly rotating objects. 
Coherent millisecond pulsations from AMSPs, indicating ongoing accretion onto the magnetic poles, place specific observational constraints on the inner disk radius, as it has to be larger than the NS surface ($\sim$10 km, depending on the equation of state), and not much larger than the corotation radius, defined as the point at which the magnetosphere rotation matches that of an assumed Keplerian disk. If the inner disk radius were significantly larger than the corotation radius, it would imply that the magnetic field is spinning faster than the disk, potentially creating a centrifugal barrier that prevents accretion. 
The corotation radius is given by $\mathrm{R_{co}} = [G \, M/(2 \pi \nu)^2]^{1/3} \simeq 1.68 \times 10^{6} \, \mathrm{M_{1.4}} \, \mathrm{P_{ms}}^{2/3} \, \mathrm{s}$, where $\mathrm{M_{1.4}}$ is the NS mass in units of 1.4 $\mathrm{M_\odot}$ and $\mathrm{P_{ms}}$ is the spin period in ms. For a 1.4 $\mathrm{M_\odot}$ NS and a spin period of $\sim$2.49 ms (Table \ref{table:Timing_sol}), we have $\mathrm{R_{co}} \simeq 31$ km. The estimated value for the inner disk radius is larger than this, but it is still compatible within the errors with the corotation radius.

\begin{table*} 
\renewcommand{\arraystretch}{1.2}
\footnotesize
\centering
\caption{List of estimated inner disk radii for AMSPs from modeling the reflection spectra.} \label{tab:Rin_estimates}
  \begin{tabular}{l c c c c}          
  \hline 
  \hline
Source & Telescope/Instrument & Model & $\mathrm{R_{in} \, (R_g)}$ & Ref.\\
\hline
SAX J1808.4$-$3658 & XMM-Newton/EPIC-pn & \texttt{diskline} & $9^{+4}_{-3}$ & [1]\\
                   & XMM-Newton/EPIC-pn + Suzaku  & \texttt{diskline} & $13.2 \pm 2.5$ & [2]\\
                   & NuSTAR & \texttt{relxillCp} & $14.9 \pm 2.5$ & [3]\\
                   & XMM-Newton/RGS, EPIC-pn & \texttt{diskline} & $10^{+8}_{-3}$ & [3]\\
IGR J17511$-$3057 & XMM-Newton/RSG, EPIC-pn & \texttt{diskline} & $37^{+31}_{-9}$ & [4]\\
                 & XMM-Newton/RSG, EPIC-pn + RXTE/PCA, HEXTE &
                 \texttt{diskline} & $22^{+9}_{-5}$ & [4]\\
                & XMM-Newton/RSG, EPIC-pn + RXTE/PCA, HEXTE & 
                 \texttt{diskline} & $27^{+6}_{-9}$ & [4]\\
HETE J1900.1$-$2455 & XMM-Newton/EPIC-pn & \texttt{rdblur*reflionx} & $23^{+10}_{-7}$  & [5]\\
                   & XMM-Newton/EPIC-pn + RXTE/PCA, HEXTE  & \texttt{rdblur*reflionx} & $25^{+16}_{-11}$  & [5]\\
SAX J1748.9$-$2021 & XMM-Newton/EPIC-pn + INTEGRAL/JEMX-1, JEMX-2, ISGRI & \texttt{diskline} & $29^{+12}_{-9}$ & [6]\\
IGR J00291$+$5934 & XMM-Newton/RGS, EPIC-pn, MOS2 + NuSTAR & \texttt{diskline} & $43^{+910}_{-18}$ & [7]\\
MAXI J1816$-$195 & NICER + NuSTAR & \texttt{relxillCp} & 5.9$\pm$0.5& [8]\\
\hline
IGR J17498$-$2921 & NICER + NuSTAR & \texttt{rdblur*rfxconv} & $21^{+17}_{-12}$& This work\\\hline
\end{tabular}
\tablecomments{References: [1]=\citet{Papitto_2009A&A}, [2]=\citet{Cackett_2009ApJ}, [3]=\citet{DiSalvo_2019MNRAS},  [4]=\citet{Papitto_2010MNRAS}, [5]=\citet{Papitto_2013MNRAS}, [6]=\citet{Pintore_2016MNRAS}, [7]=\citet{Sanna_2017MNRAS}, [8]=\citet{Li_2023MNRAS}. Uncertainties are reported at a level of statistical confidence of 90 percent.}
\end{table*}

Table \ref{tab:Rin_estimates} lists the estimated inner disk radii from reflection spectra in AMSPs. Our result falls within the range of values obtained for other sources. In general, Table \ref{tab:Rin_estimates} highlights how different models, such as \texttt{diskline}, \texttt{relxillCp}, and \texttt{rdblur*reflionx}, produce varying estimates, underscoring the complexity in fitting reflection components and the importance of advanced spectral models in achieving precise measurements. Additionally, most of the estimates in the literature are based on XMM-Newton spectra, often combined with data from other instruments. In contrast, our work and only two other studies have derived $\mathrm{R_{in}}$ estimates from NuSTAR spectra, sometimes in combination with NICER. The comparative analysis points to the need for multi-instrument observations, particularly involving XMM-Newton and NuSTAR, to further refine accretion disk radius estimates, which are crucial for testing accretion theories around quickly spinning compact objects. 

\begin{acknowledgements}
This work is based on observations acquired with the NASA mission NICER and the NuSTAR mission, which is a project led by the California Institute of Technology, managed by the Jet Propulsion Laboratory, and funded by NASA. The NuSTAR Data Analysis Software (NuSTARDAS) is jointly developed by the ASI Science Data Center (ASDC, Italy) and the California Institute of Technology (Caltech, USA). This research has made use of softwares and tools provided by the High Energy Astrophysics Science Archive Research Center (HEASARC) Online Service.
SiFAP2 observations were performed with the Italian Telescopio Nazionale Galileo (TNG) operated on the island of La Palma by the Fundación Galileo Galilei of the INAF (Istituto Nazionale di Astrofisica) at the Spanish Observatorio del Roque de los Muchachos of the Instituto de Astrofisica de Canarias.
G.I. is supported by the AASS Ph.D. joint research program between the University of Rome “Sapienza” and the University of Rome “Tor Vergata” with the collaboration of the National Institute of Astrophysics (INAF).
G.I., A.P., R.L.P, F.A., C.M. acknowledge financial support from the National Institute for Astrophysics (INAF) Research Grants ‘Uncovering the optical beat of the fastest magnetised neutron stars (FANS)’ and ‘Unveiling the secrets inside SiFAP2 data’.
G.I., A.P., R.L.P, F.A., C.M. also acknowledge funding from the Italian Ministry of University and Research (MUR) under PRIN 2020 grant No. 2020BRP57Z ‘Gravitational and Electromagnetic-wave Sources in the Universe with current and next-generation detectors (GEMS)’. A.M. is supported by the European Research Council (ERC) under the European Union’s Horizon 2020 research and innovation programme (ERC Consolidator Grant "MAGNESIA" No. 817661, PI: NR). F.C.Z. is supported by a Ram\'on y Cajal fellowship (grant agreement RYC2021-030888-I). A.M. and F.C.Z.  acknowledge support from grant SGR2021-01269 from the Catalan Government (PI: Graber/Rea).   
\end{acknowledgements}
\bibliographystyle{aa} 
\bibliography{bibliography.bib} 

\begin{thebibliography}{100}
\expandafter\ifx\csname natexlab\endcsname\relax\def\natexlab#1{#1}\fi

\bibitem[{{Alpar} {et~al.}(1982){Alpar}, {Cheng}, {Ruderman}, \& {Shaham}}]{Alpar_1982Natur}
{Alpar}, M.~A., {Cheng}, A.~F., {Ruderman}, M.~A., \& {Shaham}, J. 1982, \nat, 300, 728

\bibitem[{{Ambrosino} {et~al.}(2021){Ambrosino}, {Miraval Zanon}, {Papitto}, {Coti Zelati}, {Campana}, {D'Avanzo}, {Stella}, {Di Salvo}, {Burderi}, {Casella}, {Sanna}, {de Martino}, {Cadelano}, {Ghedina}, {Leone}, {Meddi}, {Cretaro}, {Baglio}, {Poretti}, {Mignani}, {Torres}, {Israel}, {Cecconi}, {Russell}, {Gonzalez Gomez}, {Riverol Rodriguez}, {Perez Ventura}, {Hernandez Diaz}, {San Juan}, {Bramich}, \& {Lewis}}]{Ambrosino21}
{Ambrosino}, F., {Miraval Zanon}, A., {Papitto}, A., {et~al.} 2021, Nature Astronomy, 5, 552

\bibitem[{{Ambrosino} {et~al.}(2017){Ambrosino}, {Papitto}, {Stella}, {Meddi}, {Cretaro}, {Burderi}, {Di Salvo}, {Israel}, {Ghedina}, {Di Fabrizio}, \& {Riverol}}]{Ambrosino17}
{Ambrosino}, F., {Papitto}, A., {Stella}, L., {et~al.} 2017, Nature Astronomy, 1, 854

\bibitem[{{Anitra} {et~al.}(2021){Anitra}, {Di Salvo}, {Iaria}, {Burderi}, {Gambino}, {Mazzola}, {Marino}, {Sanna}, \& {Riggio}}]{Anitra_2021A&A}
{Anitra}, A., {Di Salvo}, T., {Iaria}, R., {et~al.} 2021, \aap, 654, A160

\bibitem[{{Arnaud}(1996)}]{Arnaud_1996_XSPEC}
{Arnaud}, K.~A. 1996, in Astronomical Society of the Pacific Conference Series, Vol. 101, Astronomical Data Analysis Software and Systems V, ed. G.~H. {Jacoby} \& J.~{Barnes}, 17

\bibitem[{{Barbieri} {et~al.}(1994){Barbieri}, {Bhatia}, {Bonoli}, {Bortoletto}, {Ciani}, {Conconi}, {D'Alessandro}, {Fantinel}, {Mancini}, {Maurizio}, {Ortolani}, {Pucillo}, {Rafanelli}, {Ragazzoni}, {Zambon}, \& {Zitelli}}]{Barbieri_1994SPIE}
{Barbieri}, C., {Bhatia}, R.~K., {Bonoli}, C., {et~al.} 1994, in Society of Photo-Optical Instrumentation Engineers (SPIE) Conference Series, Vol. 2199, Advanced Technology Optical Telescopes V, ed. L.~M. {Stepp}, 10--21

\bibitem[{{Bult} {et~al.}(2022){Bult}, {Altamirano}, {Arzoumanian}, {Chakrabarty}, {Chenevez}, {Ferrara}, {Gendreau}, {Guillot}, {G{\"u}ver}, {Iwakiri}, {Jaisawal}, {Mancuso}, {Malacaria}, {Ng}, {Sanna}, {Strohmayer}, {Wadiasingh}, \& {Wolff}}]{Bult_2022ApJ}
{Bult}, P., {Altamirano}, D., {Arzoumanian}, Z., {et~al.} 2022, \apjl, 935, L32

\bibitem[{{Bult} {et~al.}(2020){Bult}, {Chakrabarty}, {Arzoumanian}, {Gendreau}, {Guillot}, {Malacaria}, {Ray}, \& {Strohmayer}}]{Bult_2020ApJ}
{Bult}, P., {Chakrabarty}, D., {Arzoumanian}, Z., {et~al.} 2020, \apj, 898, 38

\bibitem[{{Burderi} {et~al.}(2007){Burderi}, {Di Salvo}, {Lavagetto}, {Menna}, {Papitto}, {Riggio}, {Iaria}, {D'Antona}, {Robba}, \& {Stella}}]{Burderi_2007}
{Burderi}, L., {Di Salvo}, T., {Lavagetto}, G., {et~al.} 2007, \apj, 657, 961

\bibitem[{{Burderi} {et~al.}(2006){Burderi}, {Di Salvo}, {Menna}, {Riggio}, \& {Papitto}}]{Burderi_2006ApJ}
{Burderi}, L., {Di Salvo}, T., {Menna}, M.~T., {Riggio}, A., \& {Papitto}, A. 2006, \apjl, 653, L133

\bibitem[{{Cackett} {et~al.}(2009){Cackett}, {Altamirano}, {Patruno}, {Miller}, {Reynolds}, {Linares}, \& {Wijnands}}]{Cackett_2009ApJ}
{Cackett}, E.~M., {Altamirano}, D., {Patruno}, A., {et~al.} 2009, \apjl, 694, L21

\bibitem[{{Cackett} {et~al.}(2010){Cackett}, {Miller}, {Ballantyne}, {Barret}, {Bhattacharyya}, {Boutelier}, {Miller}, {Strohmayer}, \& {Wijnands}}]{Cackett_2010ApJ}
{Cackett}, E.~M., {Miller}, J.~M., {Ballantyne}, D.~R., {et~al.} 2010, \apj, 720, 205

\bibitem[{{Campana} \& {Di Salvo}(2018)}]{Campana_DiSalvo_2018ASSL}
{Campana}, S. \& {Di Salvo}, T. 2018, in Astrophysics and Space Science Library, Vol. 457, Astrophysics and Space Science Library, ed. L.~{Rezzolla}, P.~{Pizzochero}, D.~I. {Jones}, N.~{Rea}, \& I.~{Vida{\~n}a}, 149

\bibitem[{{Cardelli} {et~al.}(1989){Cardelli}, {Clayton}, \& {Mathis}}]{cardelli89}
{Cardelli}, J.~A., {Clayton}, G.~C., \& {Mathis}, J.~S. 1989, \apj, 345, 245

\bibitem[{{Chakrabarty} {et~al.}(2011){Chakrabarty}, {Markwardt}, {Linares}, \& {Jonker}}]{Chakrabarty_2011ATel}
{Chakrabarty}, D., {Markwardt}, C.~B., {Linares}, M., \& {Jonker}, P.~G. 2011, The Astronomer's Telegram, 3606, 1

\bibitem[{{Chakraborty} \& {Bhattacharyya}(2012)}]{Chakraborty_2012MNRAS}
{Chakraborty}, M. \& {Bhattacharyya}, S. 2012, \mnras, 422, 2351

\bibitem[{{Chou} {et~al.}(2008){Chou}, {Chung}, {Hu}, \& {Yang}}]{Chou_2008ApJ}
{Chou}, Y., {Chung}, Y., {Hu}, C.~P., \& {Yang}, T.~C. 2008, \apj, 678, 1316

\bibitem[{{Coti Zelati} {et~al.}(2014){Coti Zelati}, {Baglio}, {Campana}, {D'Avanzo}, {Goldoni}, {Masetti}, {Mu{\~n}oz-Darias}, {Covino}, {Fender}, {Jim{\'e}nez Bail{\'o}n}, {Ot{\'\i}-Floranes}, {Palazzi}, \& {Ram{\'o}n-Fox}}]{cotizelati14}
{Coti Zelati}, F., {Baglio}, M.~C., {Campana}, S., {et~al.} 2014, \mnras, 444, 1783

\bibitem[{{Cumming} {et~al.}(2001){Cumming}, {Zweibel}, \& {Bildsten}}]{Cumming_2001ApJ}
{Cumming}, A., {Zweibel}, E., \& {Bildsten}, L. 2001, \apj, 557, 958

\bibitem[{{Dauser} {et~al.}(2016){Dauser}, {Garc{\'\i}a}, {Walton}, {Eikmann}, {Kallman}, {McClintock}, \& {Wilms}}]{Dauser_2016A&A}
{Dauser}, T., {Garc{\'\i}a}, J., {Walton}, D.~J., {et~al.} 2016, \aap, 590, A76

\bibitem[{{Di Salvo} {et~al.}(2023){Di Salvo}, {Papitto}, {Marino}, {Iaria}, \& {Burderi}}]{DiSalvo_2023hxga.book}
{Di Salvo}, T., {Papitto}, A., {Marino}, A., {Iaria}, R., \& {Burderi}, L. 2023, in Handbook of X-ray and Gamma-ray Astrophysics (eds. C. Bambi, 147

\bibitem[{{Di Salvo} \& {Sanna}(2022)}]{DiSalvo_Sanna_2022ASSL}
{Di Salvo}, T. \& {Sanna}, A. 2022, in Astrophysics and Space Science Library, Vol. 465, Astrophysics and Space Science Library, ed. S.~{Bhattacharyya}, A.~{Papitto}, \& D.~{Bhattacharya}, 87--124

\bibitem[{{Di Salvo} {et~al.}(2019){Di Salvo}, {Sanna}, {Burderi}, {Papitto}, {Iaria}, {Gambino}, \& {Riggio}}]{DiSalvo_2019MNRAS}
{Di Salvo}, T., {Sanna}, A., {Burderi}, L., {et~al.} 2019, \mnras, 483, 767

\bibitem[{{Fabian} {et~al.}(1989){Fabian}, {Rees}, {Stella}, \& {White}}]{Fabian_1989MNRAS}
{Fabian}, A.~C., {Rees}, M.~J., {Stella}, L., \& {White}, N.~E. 1989, \mnras, 238, 729

\bibitem[{{Fabian} \& {Ross}(2010)}]{Fabian_2010SSRv}
{Fabian}, A.~C. \& {Ross}, R.~R. 2010, \ssr, 157, 167

\bibitem[{{Falanga} {et~al.}(2005){Falanga}, {Bonnet-Bidaud}, {Poutanen}, {Farinelli}, {Martocchia}, {Goldoni}, {Qu}, {Kuiper}, \& {Goldwurm}}]{Falanga_2005A&A}
{Falanga}, M., {Bonnet-Bidaud}, J.~M., {Poutanen}, J., {et~al.} 2005, \aap, 436, 647

\bibitem[{{Falanga} {et~al.}(2012){Falanga}, {Kuiper}, {Poutanen}, {Galloway}, {Bozzo}, {Goldwurm}, {Hermsen}, \& {Stella}}]{Falanga_2012A&A}
{Falanga}, M., {Kuiper}, L., {Poutanen}, J., {et~al.} 2012, \aap, 545, A26

\bibitem[{{Ferrigno} {et~al.}(2011){Ferrigno}, {Bozzo}, \& {Belloni}}]{Ferrigno_2011ATel}
{Ferrigno}, C., {Bozzo}, E., \& {Belloni}, L. G. A. P. T.~M. 2011, The Astronomer's Telegram, 3560, 1

\bibitem[{{Finger} {et~al.}(1999){Finger}, {Bildsten}, {Chakrabarty}, {Prince}, {Scott}, {Wilson}, {Wilson}, \& {Zhang}}]{Finger_1999}
{Finger}, M.~H., {Bildsten}, L., {Chakrabarty}, D., {et~al.} 1999, \apj, 517, 449

\bibitem[{{Foight} {et~al.}(2016){Foight}, {G{\"u}ver}, {{\"O}zel}, \& {Slane}}]{Foight_2016ApJ}
{Foight}, D.~R., {G{\"u}ver}, T., {{\"O}zel}, F., \& {Slane}, P.~O. 2016, \apj, 826, 66

\bibitem[{García {et~al.}(2013)García, Dauser, Reynolds, Kallman, McClintock, Wilms, \& Eikmann}]{Garcia_2013}
García, J., Dauser, T., Reynolds, C.~S., {et~al.} 2013, The Astrophysical Journal, 768, 146

\bibitem[{{Gendreau} {et~al.}(2012){Gendreau}, {Arzoumanian}, \& {Okajima}}]{NICER_Gendreau_2012}
{Gendreau}, K.~C., {Arzoumanian}, Z., \& {Okajima}, T. 2012, in Society of Photo-Optical Instrumentation Engineers (SPIE) Conference Series, Vol. 8443, Space Telescopes and Instrumentation 2012: Ultraviolet to Gamma Ray, ed. T.~{Takahashi}, S.~S. {Murray}, \& J.-W.~A. {den Herder}, 844313

\bibitem[{{Ghedina} {et~al.}(2018){Ghedina}, {Leone}, {Ambrosino}, {Meddi}, {Papitto}, {Riverol}, {Hernandez}, {Cecconi}, {Gonzalez G.}, {Perez Ventura}, \& {San Juan}}]{Ghedina_2018SPIE}
{Ghedina}, A., {Leone}, F., {Ambrosino}, F., {et~al.} 2018, in Society of Photo-Optical Instrumentation Engineers (SPIE) Conference Series, Vol. 10702, Ground-based and Airborne Instrumentation for Astronomy VII, ed. C.~J. {Evans}, L.~{Simard}, \& H.~{Takami}, 107025Q

\bibitem[{{Ghosh} \& {Lamb}(1979)}]{Ghosh_Lamb_1979ApJ}
{Ghosh}, P. \& {Lamb}, F.~K. 1979, \apj, 232, 259

\bibitem[{{Gibaud} {et~al.}(2011){Gibaud}, {Bazzano}, {Bozzo}, {Cadolle-Bel}, {Chenevez}, {Drave}, {Ferrigno}, {Gotz}, {Kadler}, {Kreykenbohm}, {Puehlhofer}, {Rodriguez}, {Sanchez-Fernandez}, \& {Watanabe}}]{Gibaud_2011ATel}
{Gibaud}, L., {Bazzano}, A., {Bozzo}, E., {et~al.} 2011, The Astronomer's Telegram, 3551, 1

\bibitem[{{Gierli{\'n}ski} {et~al.}(2002){Gierli{\'n}ski}, {Done}, \& {Barret}}]{Gierlinski_2002MNRAS}
{Gierli{\'n}ski}, M., {Done}, C., \& {Barret}, D. 2002, \mnras, 331, 141

\bibitem[{{Gierli{\'n}ski} \& {Poutanen}(2005)}]{Gierlinski_2005MNRAS}
{Gierli{\'n}ski}, M. \& {Poutanen}, J. 2005, \mnras, 359, 1261

\bibitem[{{Grebenev} {et~al.}(2023){Grebenev}, {Bryksin}, \& {Sunyaev}}]{INTEGRAL_2023ATel}
{Grebenev}, S.~A., {Bryksin}, S.~S., \& {Sunyaev}, R.~A. 2023, The Astronomer's Telegram, 15996, 1

\bibitem[{{Harrison} {et~al.}(2013){Harrison}, {Craig}, {Christensen}, {Hailey}, {Zhang}, {Boggs}, {Stern}, {Cook}, {Forster}, {Giommi}, {Grefenstette}, {Kim}, {Kitaguchi}, {Koglin}, {Madsen}, {Mao}, {Miyasaka}, {Mori}, {Perri}, {Pivovaroff}, {Puccetti}, {Rana}, {Westergaard}, {Willis}, {Zoglauer}, {An}, {Bachetti}, {Barri{\`e}re}, {Bellm}, {Bhalerao}, {Brejnholt}, {Fuerst}, {Liebe}, {Markwardt}, {Nynka}, {Vogel}, {Walton}, {Wik}, {Alexander}, {Cominsky}, {Hornschemeier}, {Hornstrup}, {Kaspi}, {Madejski}, {Matt}, {Molendi}, {Smith}, {Tomsick}, {Ajello}, {Ballantyne}, {Balokovi{\'c}}, {Barret}, {Bauer}, {Blandford}, {Brandt}, {Brenneman}, {Chiang}, {Chakrabarty}, {Chenevez}, {Comastri}, {Dufour}, {Elvis}, {Fabian}, {Farrah}, {Fryer}, {Gotthelf}, {Grindlay}, {Helfand}, {Krivonos}, {Meier}, {Miller}, {Natalucci}, {Ogle}, {Ofek}, {Ptak}, {Reynolds}, {Rigby}, {Tagliaferri}, {Thorsett}, {Treister}, \& {Urry}}]{Harrison_2013ApJ}
{Harrison}, F.~A., {Craig}, W.~W., {Christensen}, F.~E., {et~al.} 2013, \apj, 770, 103

\bibitem[{{Hartman} {et~al.}(2008){Hartman}, {Patruno}, {Chakrabarty}, {Kaplan}, {Markwardt}, {Morgan}, {Ray}, {van der Klis}, \& {Wijnands}}]{Hartman_2008ApJ}
{Hartman}, J.~M., {Patruno}, A., {Chakrabarty}, D., {et~al.} 2008, \apj, 675, 1468

\bibitem[{{Illiano} {et~al.}(2023){Illiano}, {Papitto}, {Sanna}, {Bult}, {Ambrosino}, {Miraval Zanon}, {Coti Zelati}, {Stella}, {Altamirano}, {Baglio}, {Bozzo}, {Burderi}, {de Martino}, {Di Marco}, {di Salvo}, {Ferrigno}, {Loktev}, {Marino}, {Ng}, {Pilia}, {Poutanen}, \& {Salmi}}]{Illiano_2023ApJ}
{Illiano}, G., {Papitto}, A., {Sanna}, A., {et~al.} 2023, \apjl, 942, L40

\bibitem[{{Kaastra} \& {Bleeker}(2016)}]{Kaastra_Bleeker_2016A&A}
{Kaastra}, J.~S. \& {Bleeker}, J.~A.~M. 2016, \aap, 587, A151

\bibitem[{{Kajava} {et~al.}(2011){Kajava}, {Ibragimov}, {Annala}, {Patruno}, \& {Poutanen}}]{Kajava_2011MNAS}
{Kajava}, J. J.~E., {Ibragimov}, A., {Annala}, M., {Patruno}, A., \& {Poutanen}, J. 2011, \mnras, 417, 1454

\bibitem[{{Kolehmainen} {et~al.}(2011){Kolehmainen}, {Done}, \& {D{\'\i}az Trigo}}]{Kolehmainen_2011MNRAS}
{Kolehmainen}, M., {Done}, C., \& {D{\'\i}az Trigo}, M. 2011, \mnras, 416, 311

\bibitem[{{Kulkarni} \& {Romanova}(2009)}]{Kulkarni_2009MNRAS}
{Kulkarni}, A.~K. \& {Romanova}, M.~M. 2009, \mnras, 398, 701

\bibitem[{{Kulkarni} \& {Romanova}(2013)}]{Kulkarni_2013MNRAS}
{Kulkarni}, A.~K. \& {Romanova}, M.~M. 2013, \mnras, 433, 3048

\bibitem[{{Lamb} {et~al.}(2009){Lamb}, {Boutloukos}, {Van Wassenhove}, {Chamberlain}, {Lo}, {Clare}, {Yu}, \& {Miller}}]{Lamb_2009ApJ}
{Lamb}, F.~K., {Boutloukos}, S., {Van Wassenhove}, S., {et~al.} 2009, \apj, 706, 417

\bibitem[{{Leahy} {et~al.}(1983){Leahy}, {Darbro}, {Elsner}, {Weisskopf}, {Sutherland}, {Kahn}, \& {Grindlay}}]{Leahy_1983ApJ}
{Leahy}, D.~A., {Darbro}, W., {Elsner}, R.~F., {et~al.} 1983, \apj, 266, 160

\bibitem[{{Li} {et~al.}(2023){Li}, {Tao}, {Zhang}, {Bu}, {Qu}, {Ji}, {Wang}, {Chen}, {Zhang}, {Ma}, {Yang}, {Ye}, {Zhao}, {Zhao}, {Huang}, {Ma}, {Qiao}, {Jia}, \& {Zhang}}]{Li_2023MNRAS}
{Li}, P.~P., {Tao}, L., {Zhang}, L., {et~al.} 2023, \mnras, 525, 595

\bibitem[{{Lightman} \& {Zdziarski}(1987)}]{Lightman_1987ApJ}
{Lightman}, A.~P. \& {Zdziarski}, A.~A. 1987, \apj, 319, 643

\bibitem[{{Linares} {et~al.}(2011{\natexlab{a}}){Linares}, {Altamirano}, {Watts}, {Strohmayer}, {Chakrabarty}, {Patruno}, {van der Klis}, {Wijnands}, {Casella}, {Armas-Padilla}, {Cavecchi}, {Degenaar}, {Kalamkar}, {Kaur}, {Yang}, \& {Rea}}]{Linares_2011ATel}
{Linares}, M., {Altamirano}, D., {Watts}, A., {et~al.} 2011{\natexlab{a}}, The Astronomer's Telegram, 3568, 1

\bibitem[{{Linares} {et~al.}(2011{\natexlab{b}}){Linares}, {Bozzo}, {Altamirano}, {Degenaar}, {Wijnands}, {Soleri}, {Belloni}, {Di Salvo}, {D'Ai}, {Papitto}, {Riggio}, \& {Burderi}}]{Linares_2011ATel_quiescence}
{Linares}, M., {Bozzo}, E., {Altamirano}, D., {et~al.} 2011{\natexlab{b}}, The Astronomer's Telegram, 3661, 1

\bibitem[{{Long} {et~al.}(2012){Long}, {Romanova}, \& {Lamb}}]{Long_2012NewA}
{Long}, M., {Romanova}, M.~M., \& {Lamb}, F.~K. 2012, \na, 17, 232

\bibitem[{{Lyne} \& {Graham-Smith}(1990)}]{Lyne_GrahamSmith_1990}
{Lyne}, A.~G. \& {Graham-Smith}, F. 1990, Cambridge Astrophysics Series, 16

\bibitem[{{Manca} {et~al.}(2023{\natexlab{a}}){Manca}, {Gambino}, {Sanna}, {Jaisawal}, {Di Salvo}, {Iaria}, {Mazzola}, {Marino}, {Anitra}, {Bozzo}, {Riggio}, \& {Burderi}}]{Manca_2023MNRAS_AMXP}
{Manca}, A., {Gambino}, A.~F., {Sanna}, A., {et~al.} 2023{\natexlab{a}}, \mnras, 519, 2309

\bibitem[{{Manca} {et~al.}(2023{\natexlab{b}}){Manca}, {Sanna}, {Marino}, {Di Salvo}, {Mazzola}, {Riggio}, {Deiosso}, {Cabras}, \& {Burderi}}]{Manca_2023MNRAS_LMXB}
{Manca}, A., {Sanna}, A., {Marino}, A., {et~al.} 2023{\natexlab{b}}, \mnras, 526, 1154

\bibitem[{{Marino} {et~al.}(2022){Marino}, {Anitra}, {Mazzola}, {Di Salvo}, {Sanna}, {Bult}, {Guillot}, {Mancuso}, {Ng}, {Riggio}, {Albayati}, {Altamirano}, {Arzoumanian}, {Burderi}, {Cabras}, {Chakrabarty}, {Deiosso}, {Gendreau}, {Iaria}, {Manca}, \& {Strohmayer}}]{Marino_2022MNRAS}
{Marino}, A., {Anitra}, A., {Mazzola}, S.~M., {et~al.} 2022, \mnras, 515, 3838

\bibitem[{{Marino} {et~al.}(2023){Marino}, {Russell}, {Del Santo}, {Beri}, {Sanna}, {Coti Zelati}, {Degenaar}, {Altamirano}, {Ambrosi}, {Anitra}, {Carotenuto}, {D'A{\`\i}}, {Di Salvo}, {Manca}, {Motta}, {Pinto}, {Pintore}, {Rea}, \& {van den Eijnden}}]{Marino_2023MNRAS}
{Marino}, A., {Russell}, T.~D., {Del Santo}, M., {et~al.} 2023, \mnras, 525, 2366

\bibitem[{{Molkov} {et~al.}(2024){Molkov}, {Lutovinov}, {Tsygankov}, {Suleimanov}, {Poutanen}, {Lapshov}, {Mereminskiy}, {Semena}, {Arefiev}, \& {Tkachenko}}]{Molkov_2024arXiv}
{Molkov}, S.~V., {Lutovinov}, A.~A., {Tsygankov}, S.~S., {et~al.} 2024, arXiv e-prints, arXiv:2404.19709

\bibitem[{Ng {et~al.}(2024)Ng, Ray, Sanna, Strohmayer, Papitto, Illiano, Albayati, Altamirano, Boztepe, Güver, Chakrabarty, Arzoumanian, Buisson, Ferrara, Gendreau, Guillot, Hare, Jaisawal, Malacaria, \& Wolff}]{Ng_2024}
Ng, M., Ray, P.~S., Sanna, A., {et~al.} 2024, The Astrophysical Journal Letters, 968, L7

\bibitem[{{Papitto} {et~al.}(2019){Papitto}, {Ambrosino}, {Stella}, {Torres}, {Coti Zelati}, {Ghedina}, {Meddi}, {Sanna}, {Casella}, {Dallilar}, {Eikenberry}, {Israel}, {Onori}, {Piranomonte}, {Bozzo}, {Burderi}, {Campana}, {de Martino}, {Di Salvo}, {Ferrigno}, {Rea}, {Riggio}, {Serrano}, {Veledina}, \& {Zampieri}}]{Papitto_2019ApJ}
{Papitto}, A., {Ambrosino}, F., {Stella}, L., {et~al.} 2019, \apj, 882, 104

\bibitem[{{Papitto} {et~al.}(2011{\natexlab{a}}){Papitto}, {Bozzo}, {Ferrigno}, {Belloni}, {Burderi}, {di Salvo}, {Riggio}, {D'A{\`\i}}, \& {Iaria}}]{Papitto_2011A&A}
{Papitto}, A., {Bozzo}, E., {Ferrigno}, C., {et~al.} 2011{\natexlab{a}}, \aap, 535, L4

\bibitem[{{Papitto} {et~al.}(2016){Papitto}, {Bozzo}, {Sanchez-Fernandez}, {Romano}, {Torres}, {Ferrigno}, {Kajava}, \& {Kuulkers}}]{Papitto_2016A&A}
{Papitto}, A., {Bozzo}, E., {Sanchez-Fernandez}, C., {et~al.} 2016, \aap, 596, A71

\bibitem[{{Papitto} {et~al.}(2013{\natexlab{a}}){Papitto}, {D'A{\`\i}}, {Di Salvo}, {Egron}, {Bozzo}, {Burderi}, {Iaria}, {Riggio}, \& {Menna}}]{Papitto_2013MNRAS}
{Papitto}, A., {D'A{\`\i}}, A., {Di Salvo}, T., {et~al.} 2013{\natexlab{a}}, \mnras, 429, 3411

\bibitem[{{Papitto} \& {de Martino}(2022)}]{Papitto_DeMartino_2022ASSL}
{Papitto}, A. \& {de Martino}, D. 2022, in Astrophysics and Space Science Library, Vol. 465, Astrophysics and Space Science Library, ed. S.~{Bhattacharyya}, A.~{Papitto}, \& D.~{Bhattacharya}, 157--200

\bibitem[{{Papitto} {et~al.}(2007){Papitto}, {di Salvo}, {Burderi}, {Menna}, {Lavagetto}, \& {Riggio}}]{Papitto_2007MNRAS}
{Papitto}, A., {di Salvo}, T., {Burderi}, L., {et~al.} 2007, \mnras, 375, 971

\bibitem[{{Papitto} {et~al.}(2009){Papitto}, {Di Salvo}, {D'A{\`\i}}, {Iaria}, {Burderi}, {Riggio}, {Menna}, \& {Robba}}]{Papitto_2009A&A}
{Papitto}, A., {Di Salvo}, T., {D'A{\`\i}}, A., {et~al.} 2009, \aap, 493, L39

\bibitem[{{Papitto} {et~al.}(2020){Papitto}, {Falanga}, {Hermsen}, {Mereghetti}, {Kuiper}, {Poutanen}, {Bozzo}, {Ambrosino}, {Coti Zelati}, {De Falco}, {de Martino}, {Di Salvo}, {Esposito}, {Ferrigno}, {Forot}, {G{\"o}tz}, {Gouiffes}, {Iaria}, {Laurent}, {Li}, {Li}, {Mineo}, {Moran}, {Neronov}, {Paizis}, {Rea}, {Riggio}, {Sanna}, {Savchenko}, {S{\l}owikowska}, {Shearer}, {Tiengo}, \& {Torres}}]{Papitto_2020NewAR}
{Papitto}, A., {Falanga}, M., {Hermsen}, W., {et~al.} 2020, \nar, 91, 101544

\bibitem[{{Papitto} {et~al.}(2013{\natexlab{b}}){Papitto}, {Ferrigno}, {Bozzo}, {Rea}, {Pavan}, {Burderi}, {Burgay}, {Campana}, {di Salvo}, {Falanga}, {Filipovi{\'c}}, {Freire}, {Hessels}, {Possenti}, {Ransom}, {Riggio}, {Romano}, {Sarkissian}, {Stairs}, {Stella}, {Torres}, {Wieringa}, \& {Wong}}]{Papitto_2013Natur}
{Papitto}, A., {Ferrigno}, C., {Bozzo}, E., {et~al.} 2013{\natexlab{b}}, \nat, 501, 517

\bibitem[{{Papitto} {et~al.}(2011{\natexlab{b}}){Papitto}, {Riggio}, {Burderi}, {di Salvo}, {D'A{\'\i}}, \& {Iaria}}]{Papitto_2011A&A_IGRJ00291}
{Papitto}, A., {Riggio}, A., {Burderi}, L., {et~al.} 2011{\natexlab{b}}, \aap, 528, A55

\bibitem[{{Papitto} {et~al.}(2010){Papitto}, {Riggio}, {di Salvo}, {Burderi}, {D'A{\`\i}}, {Iaria}, {Bozzo}, \& {Menna}}]{Papitto_2010MNRAS}
{Papitto}, A., {Riggio}, A., {di Salvo}, T., {et~al.} 2010, \mnras, 407, 2575

\bibitem[{{Patruno} {et~al.}(2009{\natexlab{a}}){Patruno}, {Rea}, {Altamirano}, {Linares}, {Wijnands}, \& {van der Klis}}]{Patruno_2009MNRAS}
{Patruno}, A., {Rea}, N., {Altamirano}, D., {et~al.} 2009{\natexlab{a}}, \mnras, 396, L51

\bibitem[{{Patruno} \& {Watts}(2021)}]{Patruno_Watts_2021ASSL}
{Patruno}, A. \& {Watts}, A.~L. 2021, in Astrophysics and Space Science Library, Vol. 461, Timing Neutron Stars: Pulsations, Oscillations and Explosions, ed. T.~M. {Belloni}, M.~{M{\'e}ndez}, \& C.~{Zhang}, 143--208

\bibitem[{{Patruno} {et~al.}(2009{\natexlab{b}}){Patruno}, {Wijnands}, \& {van der Klis}}]{Patruno_2009ApJ}
{Patruno}, A., {Wijnands}, R., \& {van der Klis}, M. 2009{\natexlab{b}}, \apjl, 698, L60

\bibitem[{{Pintore} {et~al.}(2016){Pintore}, {Sanna}, {Di Salvo}, {Del Santo}, {Riggio}, {D'A{\`\i}}, {Burderi}, {Scarano}, \& {Iaria}}]{Pintore_2016MNRAS}
{Pintore}, F., {Sanna}, A., {Di Salvo}, T., {et~al.} 2016, \mnras, 457, 2988

\bibitem[{{Poutanen}(2006)}]{Poutanen_2006AdSpR}
{Poutanen}, J. 2006, Advances in Space Research, 38, 2697

\bibitem[{{Protassov} {et~al.}(2002){Protassov}, {van Dyk}, {Connors}, {Kashyap}, \& {Siemiginowska}}]{Protassov_2002ApJ}
{Protassov}, R., {van Dyk}, D.~A., {Connors}, A., {Kashyap}, V.~L., \& {Siemiginowska}, A. 2002, \apj, 571, 545

\bibitem[{{Radhakrishnan} \& {Srinivasan}(1982)}]{Radhakrishnan_Srinivasan_1982CSci}
{Radhakrishnan}, V. \& {Srinivasan}, G. 1982, Current Science, 51, 1096

\bibitem[{{Riggio} {et~al.}(2011){Riggio}, {Burderi}, {di Salvo}, {Papitto}, {D'A{\`\i}}, {Iaria}, \& {Menna}}]{Riggio_2011A&A}
{Riggio}, A., {Burderi}, L., {di Salvo}, T., {et~al.} 2011, \aap, 531, A140

\bibitem[{{Riggio} {et~al.}(2008){Riggio}, {Di Salvo}, {Burderi}, {Menna}, {Papitto}, {Iaria}, \& {Lavagetto}}]{Riggio_2008Ap}
{Riggio}, A., {Di Salvo}, T., {Burderi}, L., {et~al.} 2008, \apj, 678, 1273

\bibitem[{{Romanova} {et~al.}(2004){Romanova}, {Ustyugova}, {Koldoba}, \& {Lovelace}}]{Romanova_2004ApJ}
{Romanova}, M.~M., {Ustyugova}, G.~V., {Koldoba}, A.~V., \& {Lovelace}, R.~V.~E. 2004, \apj, 610, 920

\bibitem[{{Russell} {et~al.}(2011){Russell}, {Lewis}, {Altamirano}, \& {Roche}}]{Russell_2011ATel}
{Russell}, D.~M., {Lewis}, F., {Altamirano}, D., \& {Roche}, P. 2011, The Astronomer's Telegram, 3622, 1

\bibitem[{{Sanna} {et~al.}(2018){Sanna}, {Bahramian}, {Bozzo}, {Heinke}, {Altamirano}, {Wijnands}, {Degenaar}, {Maccarone}, {Riggio}, {Di Salvo}, {Iaria}, {Burgay}, {Possenti}, {Ferrigno}, {Papitto}, {Sivakoff}, {D'Amico}, \& {Burderi}}]{Sanna_2018A&A}
{Sanna}, A., {Bahramian}, A., {Bozzo}, E., {et~al.} 2018, \aap, 610, L2

\bibitem[{{Sanna} {et~al.}(2022){Sanna}, {Bult}, {Ng}, {Ray}, {Jaisawal}, {Burderi}, {Di Salvo}, {Riggio}, {Altamirano}, {Strohmayer}, {Manca}, {Gendreau}, {Chakrabarty}, {Iwakiri}, \& {Iaria}}]{Sanna_2022MNRAS_MAXIJ1957}
{Sanna}, A., {Bult}, P., {Ng}, M., {et~al.} 2022, \mnras, 516, L76

\bibitem[{{Sanna} {et~al.}(2017{\natexlab{a}}){Sanna}, {Di Salvo}, {Burderi}, {Riggio}, {Pintore}, {Gambino}, {Iaria}, {Tailo}, {Scarano}, \& {Papitto}}]{Sanna_2017}
{Sanna}, A., {Di Salvo}, T., {Burderi}, L., {et~al.} 2017{\natexlab{a}}, \mnras, 471, 463

\bibitem[{{Sanna} {et~al.}(2023){Sanna}, {Ng}, {Guillot}, {Altamirano}, {Gendreau}, {Arzoumanian}, {Ferrara}, {Chakrabarty}, {Strohmayer}, {Ray}, {Bogdanov}, \& {Bult}}]{Sanna_2023ATel_inizio_outburst}
{Sanna}, A., {Ng}, M., {Guillot}, S., {et~al.} 2023, The Astronomer's Telegram, 15998, 1

\bibitem[{{Sanna} {et~al.}(2017{\natexlab{b}}){Sanna}, {Pintore}, {Bozzo}, {Ferrigno}, {Papitto}, {Riggio}, {Di Salvo}, {Iaria}, {D'A{\`\i}}, {Egron}, \& {Burderi}}]{Sanna_2017MNRAS}
{Sanna}, A., {Pintore}, F., {Bozzo}, E., {et~al.} 2017{\natexlab{b}}, \mnras, 466, 2910

\bibitem[{{Sharma} {et~al.}(2020){Sharma}, {Beri}, {Sanna}, \& {Dutta}}]{Sharma_2020MNRAS}
{Sharma}, R., {Beri}, A., {Sanna}, A., \& {Dutta}, A. 2020, \mnras, 492, 4361

\bibitem[{{Sharma} {et~al.}(2019){Sharma}, {Jain}, \& {Dutta}}]{Sharma_2019MNRAS}
{Sharma}, R., {Jain}, C., \& {Dutta}, A. 2019, \mnras, 482, 1634

\bibitem[{{Spitkovsky}(2006)}]{Spitkovsky_2006ApJ}
{Spitkovsky}, A. 2006, \apjl, 648, L51

\bibitem[{{Standish}(1998)}]{Standish_DE405}
{Standish}, E.~M. 1998, JPL Planetary and Lunar Ephemerides, DE405/LE405, JPL Interoffice Memo 312.F-98-048 (Pasadena, CA: NASA Jet Propulsion Laboratory)

\bibitem[{Strohmayer(2024, in prep.)}]{Strohmayer_in_prep}
Strohmayer, T.~E. 2024, in prep.

\bibitem[{{Torres} {et~al.}(2011){Torres}, {Madej}, {Jonker}, {Steeghs}, {Greiss}, {Morrell}, \& {Roth}}]{Torres_2011ATel}
{Torres}, M.~A.~P., {Madej}, O., {Jonker}, P.~G., {et~al.} 2011, The Astronomer's Telegram, 3638, 1

\bibitem[{{Vaughan} {et~al.}(1994){Vaughan}, {van der Klis}, {Wood}, {Norris}, {Hertz}, {Michelson}, {van Paradijs}, {Lewin}, {Mitsuda}, \& {Penninx}}]{Vaughan_1994}
{Vaughan}, B.~A., {van der Klis}, M., {Wood}, K.~S., {et~al.} 1994, \apj, 435, 362

\bibitem[{{Verner} {et~al.}(1996){Verner}, {Ferland}, {Korista}, \& {Yakovlev}}]{Verner_1996ApJ}
{Verner}, D.~A., {Ferland}, G.~J., {Korista}, K.~T., \& {Yakovlev}, D.~G. 1996, \apj, 465, 487

\bibitem[{{Wijnands} \& {van der Klis}(1998)}]{Wijnands_VanDerKlis_1998Natur}
{Wijnands}, R. \& {van der Klis}, M. 1998, \nat, 394, 344

\bibitem[{{Wilkinson} {et~al.}(2011){Wilkinson}, {Patruno}, {Watts}, \& {Uttley}}]{Wilkinson_2011MNRAS}
{Wilkinson}, T., {Patruno}, A., {Watts}, A., \& {Uttley}, P. 2011, \mnras, 410, 1513

\bibitem[{{Wilms} {et~al.}(2000){Wilms}, {Allen}, \& {McCray}}]{Wilms_2000ApJ}
{Wilms}, J., {Allen}, A., \& {McCray}, R. 2000, \apj, 542, 914

\bibitem[{{Zdziarski} {et~al.}(1996){Zdziarski}, {Johnson}, \& {Magdziarz}}]{Zdziarski_1996MNRAS}
{Zdziarski}, A.~A., {Johnson}, W.~N., \& {Magdziarz}, P. 1996, \mnras, 283, 193

\bibitem[{{{\.Z}ycki} {et~al.}(1999){{\.Z}ycki}, {Done}, \& {Smith}}]{Zycki_1999MNRAS}
{{\.Z}ycki}, P.~T., {Done}, C., \& {Smith}, D.~A. 1999, \mnras, 309, 561

\end{thebibliography}

\appendix
\section{Spectral analysis with the \texttt{relxillCp} model}
When fitting the reflection component in the broad-band spectrum (Sect. \ref{Sec:broad-band_spectrum}) we also tried the relativistic reflection model \texttt{relxillCp}, which employs the \texttt{nthComp} model to compute the primary source spectrum. In this model, the temperature of the seed photons for the Comptonization component is fixed at 0.05 keV and cannot be adjusted during the fitting procedure \citep[e.g.,][]{DiSalvo_2019MNRAS}. Consequently, we set the reflection fraction to negative values to account for the reflection component and added a \texttt{nthComp} component to directly represent the Comptonization continuum. The adopted model was defined as \texttt{constant*TBabs*(gaussian+nthComp+relxillCp)}. 
Given the energy resolution constraints of NuSTAR, we fixed the outer disk radius to 1000~$\mathrm{R_g}$ and employed a simple power-law relation (i.e., fixing $\mathrm{Index_1} = \mathrm{Index_2}$\footnote{\url{https://www.sternwarte.uni-erlangen.de/~dauser/research/relxill/}}; in Table \ref{tab:params_broad_band_spectrum_relxill}, we put $\mathrm{Index_1}=- \beta$ for straightforward comparison with the \texttt{rfxconv} model).

The \texttt{relxillCp} normalization is defined in the appendix of \citet{Dauser_2016A&A}. We noted that letting the \texttt{relxillCp} normalization vary would yield a reflection flux greater than the Comptonized one. 
This can be easily observed by examining the input spectrum of \texttt{relxillCp} in \texttt{XSPEC} while fixing the reflection to zero. When comparing this result to a \texttt{nthComp} model with parameters set to the same values, it was evident that the flux density predicted by \texttt{relxillCp}, without the reflection component, was $\sim$70-80 times greater than that of \texttt{nthComp}. This discrepancy is not attributable to the fixed seed photons temperature of 0.05 keV in \texttt{relxillCp}, as we verified by setting the same value in \texttt{nthComp}.

We thus tried to fix the normalization of the reflection component to match that of \texttt{nthComp} divided by a corrective factor of $\sim$73.4, which ensured that the \texttt{relxillCp} model, when devoid of reflection, reproduced the same flux of \texttt{nthComp} at 1 keV.
The best-fit spectral parameters are reported in Table \ref{tab:params_broad_band_spectrum_relxill}.
Assuming a NS mass of $1.4 \, \mathrm{M_\odot}$, the compact object's spin in the \texttt{relxill} model is $\sim$$0.16$. We verified that fixing this parameter to zero did not significantly change the fit results, hence we set $a=0$ for simplicity.
The \texttt{relxill} model returned an accretion disk density of $\log{\mathrm{N}}=18.5^{+0.4}_{-0.9}$, exceeding the model default value of $15$ \citep{Dauser_2016A&A} but aligning with the expectations for a system accreting at the rate suggested by the observed X-ray flux. Using the unabsorbed 0.5-100 keV flux, $F_X = (1.892 \pm 0.004) \times 10^{-9} \, \mathrm{erg \, cm^{-2} \, s^{-1}}$ (see Sect. \ref{sec:reflection_discussion}), and assumed a NS radius of 10 km and a distance of $\sim$7.6 kpc \citep{Linares_2011ATel}, we obtained $\dot{M} \simeq 9.8 \times 10^{16} \, \mathrm{g \, s^{-1}}$.
The electron density in the accretion columns for a fully ionized plasma is (see, e.g., Eq. (4) from \citealt{Papitto_2019ApJ}): $\mathrm{N} = \mu_e \, \rho /m_H= \mu_e \, \dot{M}/(\pi \ell^2 v_{ff} \, m_H)$,
where $\mu_e=1.18$ is the mean molecular weight per electron for a fully ionized plasma with solar abundances (i.e., with a hydrogen mass fraction of X=0.71 and a helium mass fraction of Y=0.28), and $m_H$ is the proton mass. $\rho$ is the mass density of the infalling matter given by its mass divided by the product of the columns' base area $\pi \ell^2$ (with $\ell$ being the typical accretion column transverse length scale of $\sim$5 km), and the free-fall velocity close to the NS surface $v_{ff}=\sqrt{2GM/R_{\mathrm{NS}}}$ multiplied by time. This effectively gives $\rho = \dot{M} /(\pi \ell^2 v_{ff})$.
For a 1.4 $\mathrm{M_\odot}$ NS with a 10 km radius, the accretion disk density in logarithmic units is $\log{\mathrm{N}} \sim 18.7$, consistent with our best-fit value (see Table \ref{tab:params_broad_band_spectrum_relxill}).

While the best-fit parameters obtained with the \texttt{relxillCp} model align with our findings using the \texttt{rfxconv} model (see Tables \ref{tab:params_broad_band_spectrum} and \ref{tab:params_broad_band_spectrum_relxill}), we chose not to include them in the primary analysis due to the unconstrained reflection fraction, which reached its hard limit set at $-10$.
To evaluate whether this physically unreliable parameter affected the fitting process, we calculated the relative fluxes for both the Comptonization and the reflection components. Notably, we observed a discrepancy at lower energies likely due to the difference in the seed photons' temperature between the \texttt{nthComp} model ($\sim$0.8 keV) and the fixed value of 0.05 keV in \texttt{relxillCp}. Consequently, we estimated the fluxes in the 3-79 keV range (see Table \ref{tab:params_broad_band_spectrum_relxill}). The relative flux estimates suggest that the correction factor applied to the normalization of \texttt{relxillCp} is likely appropriate. However, these results should be considered with extreme caution due to the unphysical outcome for the reflection fraction. Moreover, resolving the issue associated with the \texttt{relxillCp} normalization is beyond the scope of this work.

\begin{table} [ht!]
\renewcommand{\arraystretch}{1.2}
\centering
\caption{Best-fit parameters for the broad-band spectral model defined as \texttt{constant*TBabs*(gaussian+nthComp+relxillCp)}.} \label{tab:params_broad_band_spectrum_relxill}
  \begin{tabular}{l c c c c c}          
    \hline\hline
    Component & Parameter & Value\\
    \hline
    \scshape{Tbabs} & $\mathrm{n_H}$ ($10^{22}$ cm$^{-2}$) & $3.0 \pm 0.1$  \\
    \hline
    \scshape{Gaussian} & $\mathrm{E_{Line}} \, \mathrm{(keV)}$ & $1.66 \pm 0.05$\\
    & $\sigma \, \mathrm{(keV)}$ & $0.10^{+0.05}_{-0.04}$ \\
    & $\mathrm{Norm_{Line}}$ & $0.0015^{+0.0009}_{-0.0006}$\\
    \hline
    \scshape{nthComp} & $\Gamma$ & $2.00 \pm 0.01$\\
    & $\mathrm{kT_e} \, \mathrm{(keV)}$ & $21^{+2}_{-1}$\\
    & $\mathrm{kT_{seed}} \, \mathrm{(keV)}$ & $0.75^{+0.04}_{-0.03}$\\
    & $\mathrm{Norm_{nthComp}}$ & $0.020 \pm 0.003$\\
    & $F_\mathrm{{3-79}} \, \mathrm{(10^{-10} \, erg \, cm^{-2} \, s^{-1})}$ & $4.57^{+0.03}_{-0.05}$ \\
    \hline
    \scshape{reflection model}  &$\mathrm{\beta}$ & $-3^{+6}_{-1}$ \\
    & $\mathrm{R_{in}}$ ($\mathrm{R_g}$) & $20^{+18}_{-10}$\\
    & $\mathrm{R_{out}}$ ($\mathrm{R_g}$) & $1000^{(*)}$ \\
    & Incl. (degrees) & $72^{+10}_{-19}$\\
    & Refl. frac. & $-8^{+3}_{-2^{(\star)}}$\\
    & Redshift &  $0^{(*)}$\\
    & Fe abund. (solar units) & $1^{(*)}$ \\
    & $\log{\xi}$ & $3.3^{+0.2}_{-0.1}$\\
    & $\log{\mathrm{N}}$ ($\mathrm{cm^{-3}}$)  & $18.5^{+0.4}_{-0.9}$\\
     & $F_\mathrm{{3-79}} \, \mathrm{(10^{-10} \, erg \, cm^{-2} \, s^{-1})}$ & $2.59 \pm 0.04$ \\
    \hline
    \scshape{Total} & $\mathrm{F_{1-79}} \, \mathrm{(10^{-10} \, erg \, cm^{-2} \, s^{-1})}$ & $10.09 \pm 0.02$\\
    & $\mathrm{F_{3-79}} \, \mathrm{(10^{-10} \, erg \, cm^{-2} \, s^{-1})}$ & $7.03 \pm 0.02$\\
    \hline
    &$\chi ^2$/d.o.f & 1902.4/1857\\
    \hline
\end{tabular}
\tablecomments{$^{(*)}$ Kept frozen during the fit. 
$^{(\star)}$ The parameter reached its hard limit set at -10.
}
\end{table}
\noindent

\onecolumn
\section{Additional tables}
\begin{table}[ht!]
\renewcommand{\arraystretch}{1.4}
\centering
\caption{NICER observations of IGR J17498$-$2921 analyzed in this work to study the spectral properties of the source.}              
\label{table:NICER_obs}      
\begin{tabular}{l c c}          
\hline\hline                        
ObsID & Start Time (UTC) & Exposure (s) \\
\hline
6203770101 & 2023-04-20 00:43:20	 & 3493.0 \\
6203770102 & 2023-04-20 23:54:42	 & 3763.0 \\
6560010101 & 2023-04-21 13:49:36	 & 4458.0\\
6560010102 & 2023-04-22 09:53:15	 & 3255.0\\
6560010103 & 2023-04-23 00:14:00    & 3940.0\\
6203770103 & 2023-04-23 12:13:09	& 10996.0\\
6560010104 & 2023-04-24 00:22:00	 & 7599.0\\
6203770104 & 2023-04-24 02:09:10	 & 2823.0\\
6560010105 & 2023-04-25 00:09:20	 & 9244.0\\
6560010106 & 2023-04-26 00:36:17	 & 4372.0\\
6203770105 & 2023-04-26 03:56:00	 & 855.0 \\
6560010107 & 2023-04-27 06:01:50	 & 3367.0 \\
6560010108 & 2023-04-28 05:16:40	 & 5316.0 \\
6560010109 & 2023-04-28 23:51:45	& 5726.0 \\
6560010110 & 2023-04-30 02:11:16	 & 4304.0 \\
6560010111 & 2023-05-01 01:26:27	 & 4651.0 \\
6560010112 & 2023-05-02 02:13:45	 & 1723.0 \\
6560010113 & 2023-05-03 02:58:39	 & 1148.0 \\
\hline
\end{tabular}
\end{table}
\noindent

\setlength{\tabcolsep}{6.0pt} 
\renewcommand{\arraystretch}{1.4} 

\begin{table}
\centering
\caption{Best-fitting model continuum for each NICER observation of IGR J17498$-$2921. The associated errors are reported at 90 percent confidence level.} \label{Table:spectra_NICER_1}
\begin{tabular}{c c c c c c c c c c c}
\hline\hline
      ObsID &   $\Gamma$ &      $\mathrm{kT_{seed}}$ &           $\mathrm{Norm_{nthComp}}$ &   Unabs. 1-10 keV flux &  $\chi^{2}$/d.o.f. \\
            &            & (keV) & & $(10^{-10} \, \mathrm{erg \, s^{-1} \, cm^{-2}})$ & \\
\hline
 6203770101 & $1.79^{+0.04}_{-0.03}$ & $0.31 \pm 0.01$ & $0.134 \pm 0.004$ & $9.79 \pm 0.02$ & 110.9/119 \\
 6203770102 & $1.87^{+0.05}_{-0.04}$ & $0.35 \pm 0.01$ & $0.130 \pm 0.004$ & $10.07 \pm 0.02$ & 97.1/120  \\
 6560010101 & $1.76 \pm 0.03$ & $0.32 \pm 0.01$ & $0.131 \pm 0.003$ & $10.34 \pm 0.02$ & 125.6/121 \\
 6560010102 & $1.80 \pm 0.03$ & $0.29 \pm 0.02$ & $0.120 \pm 0.004$ & $8.37 \pm 0.02$ & 123.9/116 \\
 6560010103 & $1.72^{+0.03}_{-0.02}$ & $0.28 \pm 0.02$ & $0.103^{+0.003}_{-0.004}$ & $7.60 \pm 0.02$ & 149.6/116 \\
 6560010105 & $1.82^{+0.03}_{-0.02}$ & $0.31 \pm 0.01$ & $0.112^{+0.002}_{-0.003}$ & $8.10 \pm 0.01$ & 108.4/126 \\
 6560010106 & $1.87^{+0.03}_{-0.02}$ & $0.36 \pm 0.01$ & $0.127 \pm 0.003$ & $10.12 \pm 0.02$ & 109.4/121 \\
 6203770105 & $1.88^{+0.06}_{-0.05}$ & $0.33 \pm 0.02$ & $0.120^{+0.006}_{-0.007}$ & $8.78 \pm 0.03$ & 130.4/107 \\
 6560010107 & $1.83^{+0.04}_{-0.03}$ & $0.31^{+0.02}_{-0.01}$ & $0.105 \pm 0.003$ & $7.45 \pm 0.02$ & 96.2/116  \\
 6560010109 & $1.70 \pm 0.01$ & $0.30 \pm 0.01$ & $0.069 \pm 0.002$ & $5.31 \pm 0.01$ & 206.3/120 \\
 6560010110 & $1.72 \pm 0.02$ & $0.29^{+0.01}_{-0.02}$ & $0.064 \pm 0.002$ & $4.80 \pm 0.01$ & 190.8/118 \\
 6560010111 & $1.77 \pm 0.02$ & $0.30 \pm 0.01$ & $0.066 \pm 0.002$ & $4.69 \pm 0.01$ & 159.3/119 \\
 6560010112 & $1.91^{+0.06}_{-0.04}$ & $0.32 \pm 0.02$ & $0.078 \pm 0.004$ & $5.38 \pm 0.02$ & 101.8/108 \\
 6560010113 & $1.83 \pm 0.03$ & $0.30 \pm 0.03$ & $0.065^{+0.004}_{-0.005}$ & $4.41 \pm 0.02$ & 110.4/107 \\ 
\hline
\end{tabular}
\tablecomments{The absorption column density and the electron temperature were kept frozen at $\mathrm{n_H}=2.87 \times 10^{22} \, \mathrm{cm^{-2}}$ and $\mathrm{kT_e}=20 \, \mathrm{keV}$ during the fits.}
\end{table}

\begin{table*}
\centering
\caption{Best-fitting parameters for the emission lines detected in each NICER spectrum of IGR J17498$-$2921. 
The $\sim$1.7 keV line is most likely a Si fluorescence line from the Focal Plane Modules, while the second Gaussian line was included to account for any residual features in the energy range where the iron line is expected, i.e. $\sim$6.40-6.97 keV.
The associated errors are reported at 90 percent confidence level.} \label{Table:spectra_NICER_lines}
\begin{tabular}{c c c c c c c}
\hline\hline
        ObsID & \multicolumn{3}{c}{Si Line} & \multicolumn{3}{c}{Fe Line} \\
        &         $\mathrm{E_{Line,1}}$ &           $\sigma_\mathrm{{Line,1}}$ &            $\mathrm{Norm_{Line,1}}$            &   $\mathrm{E_{Line,2}}$ &           $\sigma_\mathrm{{Line,2}}$ &            $\mathrm{Norm_{Line,2}}$        \\
      &  (keV) & (keV) & & (keV) & (keV) &  \\
\hline
 6203770101 & $1.69^{+0.03}_{-0.04}$ & $0.08^{+0.04}_{-0.03}$   & $0.0014 \pm 0.0005$ & $6.40^{(*)}$ & $1.2^{+0.7}_{-0.5}$ & $0.001^{+0.002}_{-0.001}$ \\
 6203770102 & $1.64 \pm 0.04$        & $0.12^{+0.04}_{-0.03}$   & $0.0022^{+0.0008}_{-0.0006}$ & $6.40^{(\star)} - 6.51$   & $1.3^{+0.4}_{-0.3}$   & $0.003^{+0.002}_{-0.001}$  \\
 6560010101 & $1.69 \pm 0.03$        & $0.07 \pm 0.03$   & $0.0012 \pm 0.004$ & $6.40^{(\star)} - 6.47$   & $1.3 \pm 0.3$   & $0.002 \pm 0.001$\\
 6560010102 & $1.69 \pm 0.04$        & $0.07^{+0.04}_{-0.03}$   & $0.0010 \pm 0.0004$ & $6.40^{(\star)} - 6.54$   & $1.3 \pm 0.5$   & $0.001 \pm 0.001$ \\
 6560010103 & $1.68^{+0.05}_{-0.07}$ & $0.07 \pm 0.05$   & $0.0008^{+0.0005}_{-0.0004}$ & $6.40^{(\star)} - 6.52$   & $1.0^{+0.6}_{-0.4}$   & $0.0006^{+0.0008}_{-0.0004}$ \\
 6560010105 & $1.68 \pm 0.03$        & $0.08^{+0.03}_{-0.02}$   & $0.0014^{+0.0004}_{-0.0003}$ & $6.40^{(\star)} - 6.48$   & $1.1 \pm 0.3$   & $0.0015^{+0.0009}_{-0.0006}$ \\
 6560010106 & $1.70 \pm 0.03$        & $0.07 \pm 0.03$   & $0.0014^{+0.0005}_{-0.0004}$ & $6.40^{(\star)} - 6.54$   & $0.9 \pm 0.3$   & $0.0016^{+0.0009}_{-0.0006}$ \\
 6203770105 & $1.65^{+0.05}_{-0.06}$ & $0.09^{+0.06}_{-0.04}$   & $0.002 \pm 0.001$ & $6.40^{(\star)} - 6.56$   & $1.0 \pm 0.5$   & $0.002^{+0.002}_{-0.001}$ \\
 6560010107 & $1.70 \pm 0.04$        & $0.08^{+0.04}_{-0.03}$   & $0.0010^{+0.0004}_{-0.0003}$ & $6.40^{(\star)} - 6.62$   & $1.1^{+0.6}_{-0.5}$   & $0.001 \pm 0.001$ \\
 6560010109 & $1.7^{+0.1}_{-0.2}$    & $0.0^{(*)}$ & $0.0001 \pm 0.0001$ & - & - & - \\
 6560010110 & $1.78^{+0.03}_{-0.04}$ & $0.0^{(*)}$ & $0.0002 \pm 0.0001$ & $6.6^{+0.1}_{-0.2}$   & $0^{(*)}$ & $0.00006 \pm 0.00005$ \\
 6560010111 & $1.70 \pm 0.03$        & $0.05 \pm 0.04$   & $0.0005 \pm 0.0002$ & - & - & - \\
 6560010112 & $1.68^{+0.04}_{-0.05}$ & $0.10^{+0.05}_{-0.04}$   & $0.0013^{+0.0006}_{-0.0005}$ & $6.40^{(\star)} - 6.63$   & $1.0^{+0.6}_{-0.4}$   & $0.001 \pm 0.001$ \\
 6560010113 & $1.64^{+0.05}_{-0.07}$ & $0.10^{+0.07}_{-0.05}$   & $0.0011^{+0.0008}_{-0.0005}$ & - & - & - \\ 
\hline
\end{tabular}
\tablecomments{$^{(*)}$ Kept frozen during the fit. 
$^{(\star)}$ The parameter reached its hard limit set at 6.4 keV.
}
\end{table*}
\end{document}